\documentclass[numberedappendix]{emulateapj}
\usepackage{amssymb}

\def\deg {{\hbox{$^\circ$}}}
\def\eq#1{\begin{equation} #1 \end{equation}}

\def\Mdot {\hbox{${\dot M}_{\rm acc}$}}

\shorttitle{Environment of young stellar objects}
\shortauthors{Vinkovi\'{c} \& Jurki\'{c}}

\begin{document}

\title{Relation between the luminosity of young stellar objects and their
circumstellar environment}

\author{Dejan Vinkovi\'{c}}
\affil{Institute for Advanced Study, School of Natural Sciences,
       Einstein Drive, Princeton, NJ 08540; dejan@ias.edu}

\author{Tomislav Jurki\'{c}}
\affil{Physics Department, University of Rijeka, Omladinska 14, HR 51000 Rijeka, Croatia}

\begin{abstract}

We present a new model-independent method of comparison of
near-infrared (NIR) visibility data of young stellar objects (YSOs).
The method is based on scaling the measured baseline with the YSO's
distance and luminosity, which removes the dependence of visibility
on these two variables. We use this method to compare all available
NIR visibility data and demonstrate that it distinguishes YSOs of
luminosity $L_\star\la 10^3L_\odot$ (low-L) from YSOs of $L_\star\ga
10^3L_\odot$ (high-L).  This confirms earlier suggestions, based on
fits of image models to the visibility data, for the difference
between the NIR sizes of these two luminosity groups
\citep{Eisner04,interfer05}. When plotted against the ``scaled''
baseline, the visibility creates the following data clusters: low-L
Herbig Ae/Be stars, T Tauri stars, and high-L Herbig Be stars.  The
T Tau cluster is similar to the low-L Herbig Ae/Be cluster, which
has $\sim$7 times smaller ``scaled'' baselines than the high-L
Herbig Be cluster. We model the shape and size of clusters with
different image models and find that low-L Herbig stars are the best
explained by the uniform brightness ring and the halo model, T Tauri
stars with the halo model, and high-L Herbig stars with the
accretion disk model. However, the plausibility of each model is not
well established. Therefore, we try to build a descriptive model of
the circumstellar environment consistent with various observed
properties of YSOs. We argue that low-L YSOs have optically thick
disks with the optically thin inner dust sublimation cavity and an
optically thin dusty outflow above the inner disk regions. High-L
YSOs have optically thick accretion disks with high accretion rates
enabling gas to dominate the NIR emission over dust. Although
observations would favor such a description of YSOs, the required
dust distribution is not supported by our current understanding of
dust dynamics.

\end{abstract}


\section{Introduction}

Young stellar objects (YSOs) are surrounded by dust and gas leftovers
from the process of star formation. Observations have revealed that
this material concentrates into a (protoplanetary) disk accreting
toward the central star. The disk undergoes the process of dust and
gas coagulation and formation of larger objects, eventually resulting
in a planetary system. Hence, YSOs are of a topical astronomical
interest because they can help us understand how planetary systems
form and evolve.

Dust in the disk is very efficient in absorbing the stellar radiation. It makes
the disk appear optically thick until the dust is either removed or
coagulated into larger objects. Dust is also a very efficient tracer
of protoplanetary disks in infrared, where dust reemits the energy
that it absorbed from the star. Direct imaging has proved to be
especially useful in unraveling properties of YSOs. Depending on the
wavelength, images capture different temperature zones and optical
depths of the protoplanetary dusty disk. From the terrestrial planet
formation perspective, the most interesting is the inner few AU of the
disk. In this zone dust temperatures can reach a thousand degrees or
more. At these temperatures dust either sublimates or becomes heavily
thermally processed, and it emits in the near-infrared (NIR).
Unfortunately, the angular size of this zone is on the miliarcseconds
scale and until recently unresolvable.

Thanks to advances in NIR interferometry \citep[e.g.][]{Monnier03}, the inner
regions of many YSO disks have been resolved to date. One of the first
surprises coming out of these observations is a significantly larger
NIR size than previously predicted from theoretical models of
accretion disks \citep{interfer,Tuthill01}. This led to revisions of
the existing theories of disk structure by invoking an optically thin
inner disk hole around the star. The hole is clear of dust because of
its sublimation, while the gas is unable to provide a considerable
optical thickness. On the other hand, the spectrum shows that the
inner disk emits much more NIR flux than expected from a simple flat
disk. This directly implies that the disk structure has to flare up in
the zone where it reaches its dust sublimation temperature
\hbox{($T_s\sim$1500-2000K)} in order to increase its NIR emitting area.
Such a model was described by \citet{DDN}, where the disk is
vertically puffed up at the inner disk rim because of the direct
stellar heating of the disk interior.

Unfortunately, the NIR interferometric observations were capable of
providing only the characteristic size of the emitting region, but they
were not good enough to constrain the exact geometry of this
region. It is only the most recent observations that manage to make a
step further and detect deviations from simple centrally symmetric
images \citep{skew}. These deviations turned out to be surprisingly
small, unlike strong brightness asymmetries predicted by inclined disk
models of \citet{DDN}. Such a result favors a derivative of this model
where the inner disk rim does not create a vertically flat wall, but
rather a curved surface \citep{Isella}. In addition, it is still not
clear from the data if an additional dusty wind component coexists
with the disk and contributes to the images and NIR flux. According to
\citet{VIJE06}, such an optically thin halo around the inner disk can
completely explain the NIR flux without a need for the disk puffing.

Another major result coming out of the NIR interferometric data is the
dependence of inner disk size on luminosity
\citep{interfer,interfer05}.  The size was derived by fitting ad hoc
uniform brightness ring image models to the measured visibilities. The
inner ring radius $R_{in}$, which is assumed to be the dust
sublimation radius of the disk, is plotted against the luminosity
$L_\star$ of the observed object. The expected trend $R_{in}\propto
L_\star^{1/2}$ is followed by YSOs of $L_\star<10^3L_\odot$, with sublimation
temperatures between 1000 K and 1500 K. Objects with $L_\star>10^3L_\odot$
deviate from this trend by having considerably smaller inner radius
than expected. \citet{interfer05} argue that high luminosity objects
have optically thick gas within the inner dust-free disk hole. This
would shield the dust from direct stellar heating and enable the dust
to survive closer to the star, resulting with a smaller inner ring
radius.

The basic approach in these studies is to use predefined theoretical
models and correlate the object's luminosity with a model parameter
derived from fitting the model to the visibility data.  This
inevitably leads to questions on the validity of the model and
conclusions derived from it. This is especially true for models of
inner disk, considering large uncertainties about the inner disk
geometry and dust properties. In this paper we propose an approach
that can detect visibility dependence on luminosity without invoking any
model of dust geometry. The approach is based on scaling the baseline
value such that all objects seemingly appear located at the same
distance from us and have the same luminosity.

We collected all available NIR visibility data on YSOs (13 T Tau, 27
Herbig Ae/Be and 4 FU Ori stars) in the literature and in
\S\ref{section_data} we compared their scaled visibilities. We confirm
that the circumstellar environment of objects with
$L_\star>10^3L_\odot$ differs from less luminous objects, with high
luminous objects showing smaller structures. After that in
\S\ref{section_theory_models} we plot scaled theoretical visibilities
of uniform brightness ring, dusty halo, and accretion disk over the
scaled measured visibilities and explore the range of model parameters
that are needed for accommodating all the data. In
\S\ref{section_doscussion} we present an extensive discussion of
various observational aspects of YSOs, with a special attention given
to the differences between low- and high-luminous YSOs. We try to
sketch a self-consistent model that would explain various observed
properties of YSOs. The summary of the paper is given in
\S\ref{section_summary}.\\

\begin{figure*}
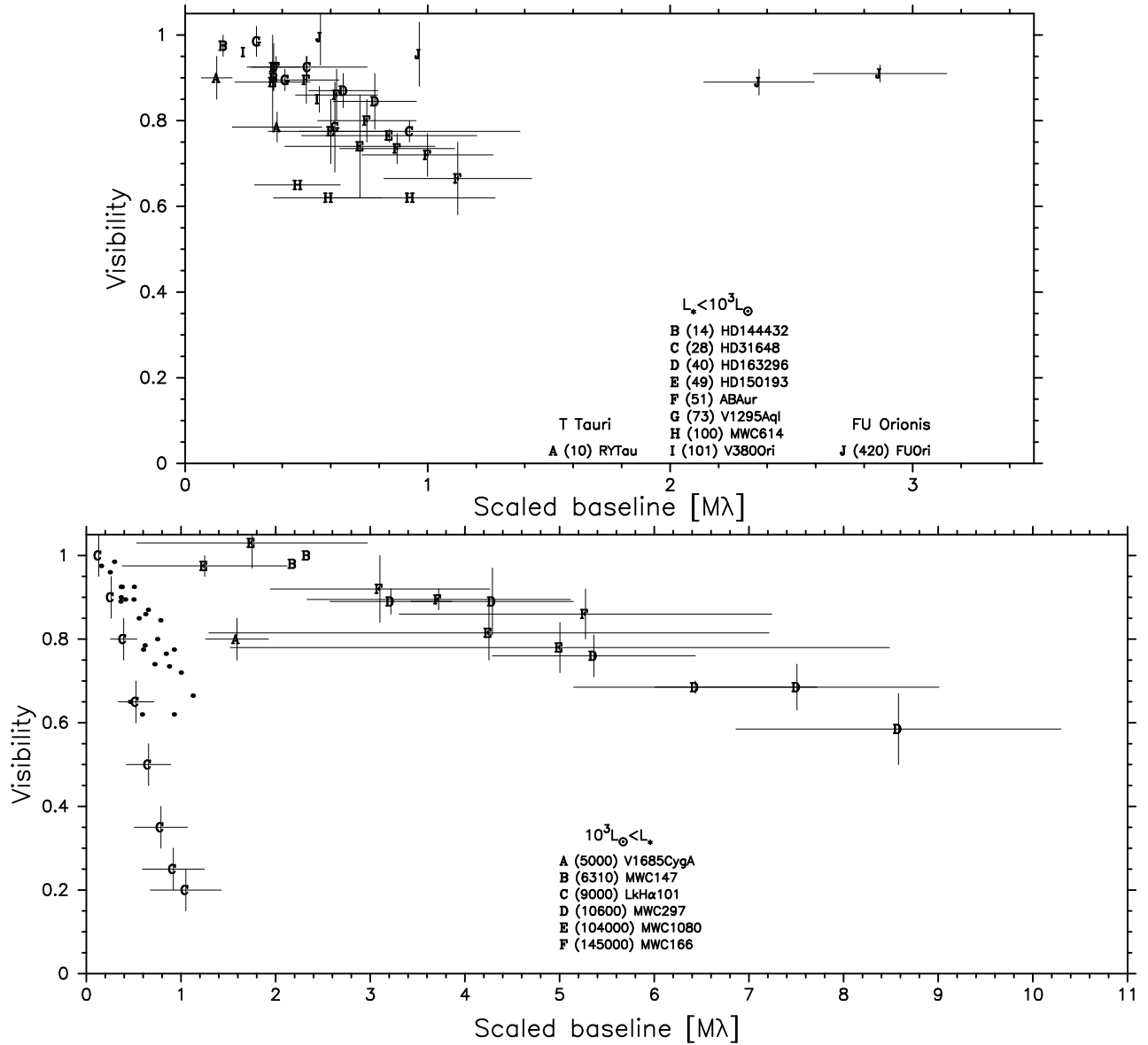

 \begin{center}
  \includegraphics[width=7.8cm,angle=-90]{Figure_1a.ps}\\
  \includegraphics[width=7.8cm,angle=-90]{Figure_1b.ps}
 \caption{\label{H-band-data}
H-band visibility data with scaled baseline (see equation
\ref{Bscaled}). Letters mark the averaged visibility data from Table
\ref{table-data}. The vertical error bar is a scatter of measured
visibility values, while the horizontal error bar derives from
uncertainties in distance and luminosity. The upper panel shows the
data for a T Tauri star, Herbig Ae/Be stars of $L_\star<10^3L_\odot$ and FU
Orionis. The lower panel shows Herbig Be stars of
$L_\star>10^3L_\odot$. Notice the tendency of low luminosity stars (except
for FU Ori) to cluster very compactly, which is a signature of very
similar images of diffuse radiation, while high luminosity
stars have relatively small circumstellar structures.  For
comparison, dots in the lower panel show the location of low luminous
stars. Stellar luminosities in units of $L_\odot$ are indicated
together with the stellar name.}
 \end{center}
\end{figure*}

\begin{figure*}
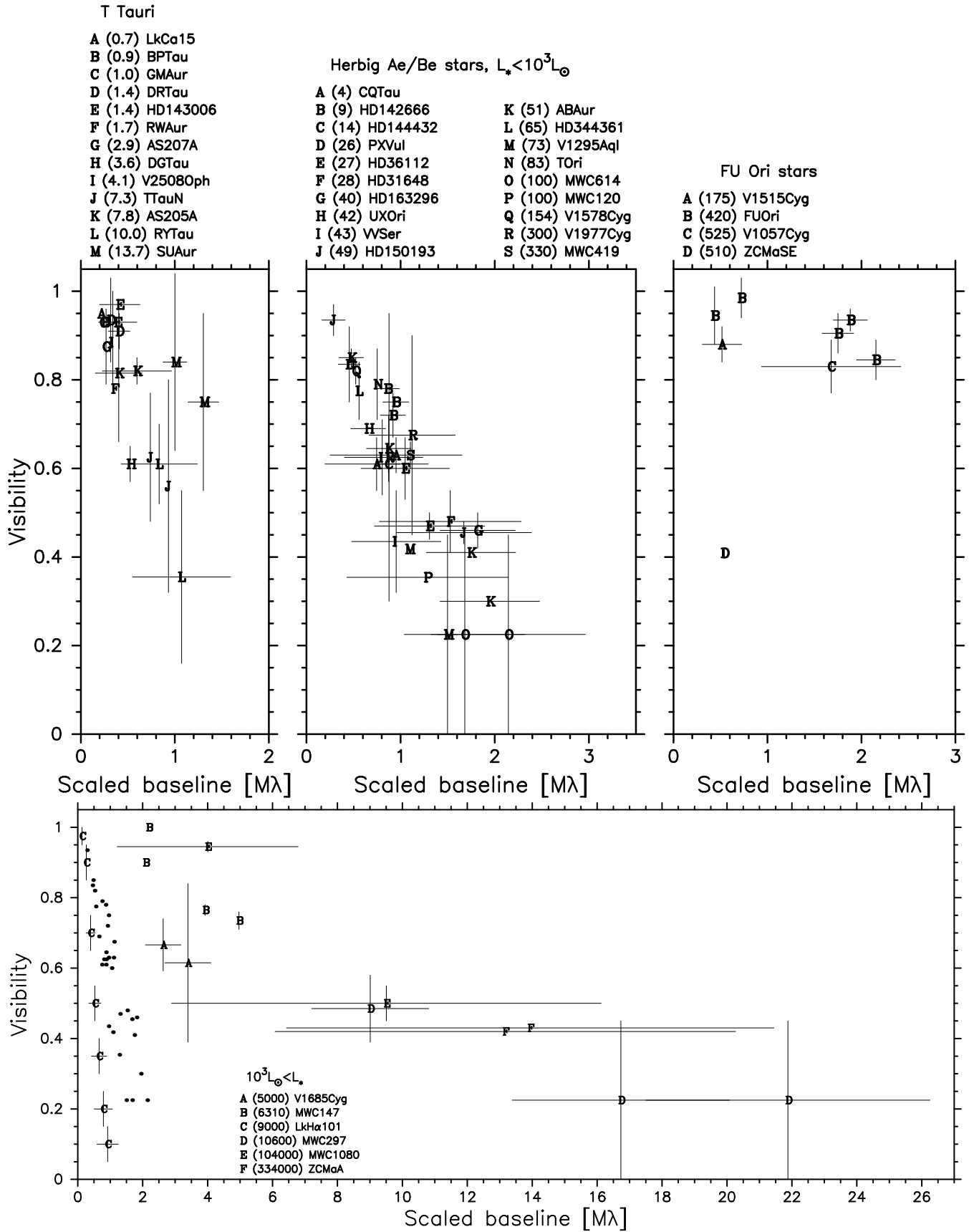

 \begin{center}
 \includegraphics[width=14.8cm,angle=-90]{Figure_2a.ps}\\
 \includegraphics[width=8cm,angle=-90]{Figure_2b.ps}
 \caption{\label{K-band-data}
Same as Figure \ref{H-band-data}, but for K-band data.  T Tauri,
Herbig Ae/Be and FU Orionis are now separated for clarity into their
own panels. The clustering of low luminosity objects and their
distinction from high luminosity objects is clearly visible. The
clustering is even more evident as we parametrize the data by various
visibility functions (see Figures
\ref{fig_inclinations}-\ref{fig_acc_disk_FUOri}).}
 \end{center}
\end{figure*}

\section{YSO visibilities with scaled baselines}
\label{section_data}

Studies of the relation between YSO luminosities and their NIR
visibilities have been based on fitting the constant surface
brightness ring image model to the visibility data. The obtained inner
ring radius is then correlated with the luminosity, where the basic
assumption is that this radius is equivalent to the dust sublimation
distance from the star
\citep{interfer,Eisner04,interfer05,Akeson05a}. A slightly different
approach is used by \citet{VIJE06} who use a dusty halo model, but the
approach is still based on obtaining the sublimation radius from the
fit to visibility data. Using this approach, \citet{Eisner04} and
\citet{interfer05} showed that YSOs with luminosity
$\ga$10$^3L_\odot$ have a significantly smaller inner ring radius than
less luminous objects.

It is, however, not clear if the ring model is a plausible model for
fitting the visibility. The NIR disk surface brightness is not a
uniform ring, but a complicated function that critically depends on
dust emission and scattering properties and the disk optical depth
structure. Moreover, \citet{VIJE06} used a halo model and obtained a
significantly smaller scatter of inner halo radius with luminosity.
Here we show that, when exploring correlations between luminosity
and visibility data, all uncertainties due to different models of
circumstellar dust properties and geometry can be avoided by model
independent scaling of visibility.

The visibility $V(B)$ is a function of the spatial frequency $B$, also
called baseline. It consists of two individual additive visibilities:
stellar and diffuse. The stellar contribution to the NIR visibility at
currently used baselines is a constant because the star remains
unresolved. The NIR diffuse part is a contribution from the
circumstellar dust emission and scattering, while the gas brightness
can be neglected (we will see later on that this is not true for
luminous YSOs).

The diffuse visibility critically depends on two parameters: i)
distance to the object, which affects the angular size of the image on
the sky, and ii) physical size of the object, which is controlled by
dust radiative transfer. The equations of dust radiative transfer have
powerful intrinsic scaling properties according to which luminosity
and linear dimensions are irrelevant for solving the equations
\citep{IE97}. Overall luminosity is never an input parameter: only the
source spectral shape is important. Also, absolute scales of dust
densities and distances are irrelevant: only the optical depth,
geometrical angles, relative thicknesses and aspect ratios enter
the equations. Luminosity is important only when we need to translate
dimensionless radiative transfer solutions to physical units,
where physical dimensions scale with $\sqrt{L_\star}$ \citep[equation 27
in][]{IE97}.

When comparing intrinsic properties of the circumstellar dust geometry
of two objects in the sky, we want to make sure that detected size
differences are not due to: a) differences in distances, and b) size
scaling due to different luminosities.  Therefore, we first need to
scale objects to the same luminosity and to the same
distance. Visibility of an image is inversely proportional to the
characteristic angular size of the image \citep{IE96}. In turn, the
angular size is inversely proportional to distance $D$, but, as we
argue from the radiative transfer scaling properties, proportional to
$\sqrt{L_\star}$. Hence, we define the {\it scaled baseline}
\eq{\label{Bscaled}
 B_{\rm scaled} = B\cdot {\sqrt{L_\star/L_\odot}\over D/{\rm pc}},
}
which has the same physical units like the ordinary baseline, but it
removes the intrinsic dependence on distance and radiative transfer scaling.

Although this scaling can be postulated for any system, it is
meaningful only for systems with a central heating source. Binary
systems, for example, can have entirely different scaling
properties. Notice, however, that the method is useful for disks
with significant accretion luminosities because the most of
accretion luminosity comes from the accretion shock. Hence, the dust
in such disks is exposed primarily to the sum of stellar and
accretion luminosity $L_\star+L_{\rm acc}$ and this sum should
replace the luminosity factor in equation \ref{Bscaled}.

We collected all objects with available H- and K-band visibility data
and list them in Table \ref{table-data}: 13 objects classified as T
Tauri stars, 27 Herbig Ae/Be stars and 4 FU Orionis objects.  We
excluded from the list objects that are most probably B[e] supergiants
(MWC 349A, MWC342, HD45677, HD58647) and show them separately only for
illustration. Since the goal is to compare overall trends in
visibility, we average the visibility data within bins of about
0.5$M\lambda$ in objects that have many measurements taken at similar
baselines (the lower and upper visibility limits are shown in the
table when applicable). For a more detailed view of the visibility
data in individual objects, we suggest following the references given
in the table.

Figures \ref{H-band-data} and \ref{K-band-data} show the visibility
data when plotted against the scaled baseline. The difference between
low luminosity ($L_\star\la 10^3 L_\odot$, low-L) and high luminosity
($L_\star\ga 10^3 L_\odot$, high-L) Hebig Ae/Be objects is now clearly
visible. It is important to emphasize that this is a {\it model
independent} detection of intrinsic differences between the
circumstellar environments of low-L and high-L objects. Clustering is
evident in low-L Herbig Ae/Be stars and T Tau stars, while high-L
objects cluster at significantly larger scaled baselines (except for
LkH$\alpha$ 101). The only exception are FU Ori objects, which appear
more similar to high-L Herbig Be stars than to their older
counterparts - T Tau stars. This suggests possible similarities in
circumstellar geometry between these two types of objects.  We address
this issue in \S\ref{HerbigBe_disk_nature}.

Figure \ref{K-band-data} indicates a slight difference between
zones of T Tau clustering and low-L Herbig Ae/Be clustering. This is a
result of differences in the stellar spectral shape, where T Tau stars
contribute relatively more to the NIR flux then Herbig Ae/Be stars.
How that influences the scaled visibility curves will be more apparent
in the next section when we plot theoretical models over the data.

Trends in Figures \ref{H-band-data} and \ref{K-band-data} are so
strong that we can use them to identify two anomalous individual
cases:

\begin{deluxetable*}{lcccccc}
\tabletypesize{\scriptsize}
\tablecaption{\label{table-data}
Objects properties and averaged visibility data}
\tablewidth{0pt}
\tablehead{
\colhead{} &
\colhead{} &
\colhead{} &
\multicolumn{4}{c}{Visibility Averages} \\
\cline{4-7} \\
\colhead{Object Names} &
\colhead{Distance (pc)} &
\colhead{Luminosity ($L_\odot$)} &
\colhead{Baseline (M$\lambda$)} &
\colhead{H band} &
\colhead{K band} &
\colhead{Reference(s)}
}

\startdata
          &               &              &      &           &           & \\
\multicolumn{7}{c}{T Tau OBJECTS} \\
          &               &              &      &           &           & \\
LkCa 15  & 142$\pm$14 (1)  & 0.74 (2)        & 35.5 &     & 0.95      & 2\\
BP Tau   & 142$\pm$14 (1)  & 0.83 (2)        & 37.5 &     & 0.93      & 2\\
DR Tau   & 142$\pm$14 (1)  & 1.28$\pm$0.50 (51)& 38.0 &     & 0.84-1.03 & 3\\
         &                  &                  & 48.9 &     & 0.87-0.95 & \\
GM Aur   & 142$\pm$14 (1)  & 1.01 (2)        & 33.3 &     & 0.93      & 2\\
AS 205A  & 165$\pm$20 (5)  & 7.8$\pm$6.5 (4) & 25.0 &     & 0.66-0.97 & 4\\
         &                  &                  & 37.0 &     & 0.79-0.85 & \\
HD143006 & 94$\pm$35 (6)   & 1.4$\pm$0.5 (6) & 25.5 &     & 0.93      & 6\\
         &                  &                  & 26.9 &     & 0.97      & \\
RW Aur   & 142$\pm$14 (1)  & 1.7 (2)         & 39.0 &     & 0.78      & 2\\
AS 207A  & 165$\pm$20 (5)  & 2.9$\pm$0.2 (4) & 25.5 &     & 0.79-0.96 & 4\\
V2508 Oph & 165$\pm$20 (5) & 4.1$\pm$0.8 (4) & 27.0 &     & 0.77-1.00 & 4\\
DG Tau   & 142$\pm$14 (1)  & 3.62 (2)        & 38.8 &     & 0.57-0.65 & 2,7\\
T Tau N  & 142$\pm$14 (1)  & 7.3 (52)         & 38.5 &     & 0.48-0.77 & 3,8,9\\
         &                  &                  & 48.5 &     & 0.32-0.80 & \\
RY Tau   & 142$\pm$14 (1)  & 10$\pm$7(10,51)  & 6.0  & 0.85-0.95 &     & 3,10\\
         &                  &                  & 17.5 & 0.75-0.82 &     & \\
         &                  &                  & 38.5 &     & 0.52-0.70 & \\
         &                  &                  & 49.5 &     & 0.16-0.55 & \\
SU Aur   & 142$\pm$14 (1)  & 13.5$\pm$0.7 (51)        & 38.0 &     & 0.64-1.04 & 3,8,9\\
         &                  &                  & 49.5 &     & 0.55-0.95 & \\
          &               &              &      &           &           & \\
\cline{1-7} \\
\multicolumn{7}{c}{Herbig Ae/Be OBJECTS} \\
          &               &              &      &           &           & \\
PX Vul   & 420        (11) & 26.2$\pm$12.5 (4) & 38.5 &   & 0.75-0.92 & 4\\
CQ Tau   & 104$\pm$21 (12) & 4.2$\pm$3.8 (12,45) & 38.0 &  & 0.55-0.67 & 13\\
         &                  &                  & 48.5 &     & 0.59-0.67 & \\
HD142666 & 116 (14)        & 8.8$\pm$2.5 (6) & 34.1 &    & 0.78      & 6\\
         &                  &                  & 36.3 &     & 0.67-0.77 & \\
         &                  &                  & 37.5 &     & 0.75      & \\
HD144432 & 145 (15)  & 14.5$\pm$4.0 (10,6) & 6.0  & 0.95-1.00 &     & 10,6\\
         &                  &                  & 14.0 & 0.87-0.93 &     & \\
         &                  &                  & 33.5 &     & 0.59-0.63 & \\
HD36112 (MWC758) & 210$\pm$50 (12) & 27$\pm$12 (12,6) & 38.9 &  & 0.53-0.67 & 13,6\\
         &                  &                  & 48.5 &     & 0.44-0.50 & \\
HD163296 (MWC275) & 124$\pm$15 (12) & 40$\pm$8 (10,6) & 7.0 & 0.87-0.98 & & 10,6,16\\
         &                  &                  & 12.5 & 0.83-0.91 &     & \\
         &                  &                  & 15.0 & 0.78-0.91 &     & \\
         &                  &                  & 34.9 &      & 0.42-0.50 & \\
UX Ori   & 400$\pm$60 (18) & 42.5$\pm$11.5 (17) & 38.9 &    & 0.69  & 6\\
HD31648 (MWC480) & 134$\pm$21 (12) & 28$\pm$18 (10,12) & 12.5 & 0.90-0.95 & & 10,13\\
         &                  &                  & 23.0 & 0.75-0.80 &     & \\
         &                  &                  & 38.0 &      & 0.41-0.55 & \\
HD150193 (MWC863) & 134$\pm$21 (12) & 28$\pm$18 (6,10) & 6.0 & & 0.90-0.97 & 6,10,16\\
         &                  &                  & 7.5 & 0.78-1.00 & & \\
         &                  &                  & 12.5 & 0.70-0.85 & & \\
         &                  &                  & 15.0 & 0.62-0.86 & & \\
         &                  &                  & 17.5 & 0.75-0.78 & & \\
         &                  &                  & 34.8 & & 0.43-0.48 & \\
AB Aur   & 147$\pm$20 (12) & 51$\pm$14 (12)  & 7.5  & 0.90-0.95 & & 16,13,19,10,20\\
         &                  &                  & 9.6  & & 0.83-0.87 & \\
         &                  &                  & 10.0 & 0.84-0.95 & & \\
         &                  &                  & 12.5 & 0.80-0.92 & & \\
         &                  &                  & 15.0 & 0.75-0.85 & & \\
         &                  &                  & 17.5 & 0.70-0.77 & 0.57-0.72 & \\
         &                  &                  & 20.0 & 0.67-0.77 & & \\
         &                  &                  & 22.5 & 0.58-0.75 & & \\
         &                  &                  & 35.0 & & 0.41 & \\
         &                  &                  & 39.0 & & 0.3  & \\
HD344361 (WW Vul) & 550 (21) & 65$\pm$5 (18) & 38.2 & & 0.71-0.84 & 6\\
V1295 Aql (MWC325, & 290$^a$ (12) & 73$\pm$10 (10,12,22)
                                               & 10.0 & 0.95-1.02 & & 16,13,10,23\\
\,\,\,\,\,\,\,\,\, HD190073) &     &                  & 14.0 & 0.87-0.92 & & \\
         &                  &                  & 21.0 & 0.68-0.89 & & \\
         &                  &                  & 37.0 & & 0.42 & \\
         &                  &                  & 51.0 & & 0.00-0.45 & \\
T Ori    & 460 (18)        & 83 (18)         & 38.0 & & 0.71-0.87 & 13\\
VV Ser    & 350$\pm$100 (18,24,25,26)  & 43$\pm$20 (18,27)
                                         & 38.0 & & 0.54-0.71 & 13,19\\
          &               &              & 45.0 & & 0.32-0.55 & \\
\enddata
\end{deluxetable*}

\setcounter{table}{0}
\begin{deluxetable*}{lcccccc}
\tabletypesize{\scriptsize}
\tablecaption{Continued...}
\tablewidth{0pt}
\tablehead{
\colhead{} &
\colhead{} &
\colhead{} &
\multicolumn{4}{c}{Visibility Averages} \\
\cline{4-7} \\
\colhead{Object Names} &
\colhead{Distance (pc)} &
\colhead{Luminosity ($L_\odot$)} &
\colhead{Baseline (M$\lambda$)} &
\colhead{H band} &
\colhead{K band} &
\colhead{Reference(s)}
}

\startdata
V1578Cyg (AS477) & 900 (28) & 154$\pm$20 (18)  & 38.1 & & 0.79-0.85 & 6\\
V1977Cyg (AS442) & 700 (29) & 300$\pm$210 (6)  & 38.3 & & 0.30-0.95 & 6,13,19\\
          &               &              & 49.0 & & 0.45-0.90 & \\
MWC419 (V594 Cas) & 650 (18) & 330 (18)  & 39.5 & & 0.60-0.66 & 23\\
MWC614 (HD179218) & 255$\pm$55 (12) & 100$\pm$35$^b$ (10)
                                         & 11.0 & 0.65 & & 16,23,10\\
          &               &              & 14.0 & 0.62 & & \\
          &               &              & 22.0 & 0.62 & & \\
          &               &              & 40.0 & & 0.0-0.45 & \\
          &               &              & 51.0 & & 0.0-0.45 & \\
MWC120 (HD37806) & 360$\pm$130 (12,30) & 100$\pm$68 (12,30)  & 38.0 & & 0.35 & 13\\
LkH$\alpha$ 101 & 340$^c$ (31,32) & 9000$\pm$5700$^c$ (31,32)  & 0.5 & 0.95-1.05 & 0.95-1.0 & 32\\
          &               &              & 1.0 & 0.85-0.95  & 0.85-0.95 & \\
          &               &              & 1.5 & 0.75-0.85  & 0.65-0.75 & \\
          &               &              & 2.0 & 0.60-0.70  & 0.45-0.55 & \\
          &               &              & 2.5 & 0.45-0.55  & 0.30-0.40 & \\
          &               &              & 3.0 & 0.30-0.40  & 0.15-0.25 & \\
          &               &              & 3.5 & 0.20-0.30  & 0.05-0.15 & \\
          &               &              & 4.0 & 0.15-0.25  &           & \\
V380 Ori  & 445$\pm$15 (18,30) & 101$\pm$16 (18,33) & 11.0 & 0.96 & & 16\\
          &               &              & 24.5 & 0.82-0.88 &           & \\
MWC147 (HD259431) & 800 (18) & 6300 (18) & 21.0 &         & 0.90      & 16,8,23\\
          &               &              & 22.0 & 0.98      & 1.00      & \\
          &               &              & 23.5 & 1.00      &           & \\
          &               &              & 39.5 &           & 0.75-0.78 & \\
          &               &              & 50.0 &           & 0.71-0.76 & \\
V1685Cyg (MWC340, & 1000 (18,19,34,35,36) & 5000$\pm$2000 (18,37)
                                         & 23.0 & 0.75-0.85 &           & 6,13,16,19\\
\,\,\,\,\,\,\,\,\, BD+40\deg 4124) & &       & 38.0 &           & 0.59-0.74 & \\
          &               &              & 49.0 &           & 0.39-0.84 & \\
MWC297 (NZ Ser) & 250$\pm$50 (38) & 33000$^d$$\pm$13000 (10,18,38)
                                         & 7.5  & 0.86-0.92 &           & 10,13,16,23,39\\
          &               &              & 10.0 & 0.81-0.97 &           & \\
          &               &              & 12.5 & 0.71-0.81 &           & \\
          &               &              & 15.0 & 0.67-0.70 &           & \\
          &               &              & 17.5 & 0.63-0.74 &           & \\
          &               &              & 20.0 & 0.50-0.67 &           & \\
          &               &              & 21.0 &           & 0.39-0.58 & \\
          &               &              & 39.0 &           & 0.00-0.45 & \\
          &               &              & 51.0 &           & 0.00-0.45 & \\
MWC1080   & 1600$\pm$600 (18,40) & 104000$\pm$76000 (10,18,27)
                                         & 5.0  & 0.95-1.00 &           & 10,13,16,19\\
          &               &              & 7.0  & 0.97-1.09 &           & \\
          &               &              & 16.0 &           & 0.93-0.96 & \\
          &               &              & 17.0 & 0.75-0.88 &           & \\
          &               &              & 20.0 & 0.72-0.84 &           & \\
          &               &              & 38.0 &           & 0.45-0.55 & \\
MWC166 (HD53367) &  1150 (18) & 145000$\pm$95000 (18,10,41)
                                         & 10.0 & 0.84-1.00 &           & 10,16\\
          &               &              & 12.0 & 0.87-0.92 &           & \\
          &               &              & 17.0 & 0.80-0.92 &           & \\
Z CMa  A  & 1100$\pm$50 (11,43,44) & 333000$\pm$268000 (42)
                                         & 27.4 &           & 0.42      & 6\\
          &               &              & 29.0 &           & 0.43      & \\
          &               &              &      &           &           & \\
\cline{1-7} \\
\multicolumn{7}{c}{FU Ori OBJECTS$^e$} \\
          &               &              &      &           &           & \\
FU Ori    & 450 (44) & 420$\pm$80 (44,46) & 9.6 &        & 0.88-1.01 & 47,48\\
          &               &              & 12.3 & 0.93-1.06 &           & \\
          &               &              & 15.9 &           & 0.94-1.03 & \\
          &               &              & 21.3 & 0.88-1.03 &           & \\
          &               &              & 38.6 &           & 0.86-0.95 & \\
          &               &              & 41.5 &           & 0.91-0.96 & \\
          &               &              & 47.5 &           & 0.80-0.89 & \\
          &               &              & 52.2 & 0.86-0.92 &           & \\
          &               &              & 63.2 & 0.89-0.93 &           & \\
V1515 Cyg   & 1000$\pm$200 (44) & 175$\pm$75 (44,46)
                                         & 36.7 &        & 0.84-0.92 & 49\\
V1057 Cyg   & 550$\pm$100 (50) & 525$\pm$275 (44,46)
                                         & 38.5 &        & 0.77-0.89 & 23,49\\
ZCMa  SE    & 1100$\pm$50 (11,43,44) & 510$\pm$90 (44,46)
                                         & 26.3 &        & 0.40-0.42 & 49\\
\enddata
\end{deluxetable*}
\setcounter{table}{0}
\begin{deluxetable}{lcccccc}[t]
\tabletypesize{\scriptsize}
\tablecaption{Continued...}
\tablewidth{250pt}
\tablehead{
}

 \tablecomments{
References: (1) \citet{Wichman}, (2) \citet{Akeson05a}, (3)
\citet{Akeson05b}, (4) \citet{Eisner05}, (5) \citet{Chini}, (6)
\citet{interfer05}, (7) \citet{Colavita}, (8) \citet{Akeson00}, (9)
\citet{Akeson02}, (10) \citet{skew}, (11) \citet{Herbst}, (12)
\citet{Ancker98}, (13) \citet{Eisner04}, (14) \citet{Meeus}, (15)
\citet{Perez}, (16) \citet{MG01}, (17) \citet{Hernandez}, (18)
\citet{Hillenbrand}, (19) \citet{Eisner03}, (20) \citet{MG99}, (21)
\citet{Friedemann}, (22) \citet{Acke05}, (23) \citet{Wilkin}, (24)
\citet{Chavarria}, (25) \citet{deLara}, (26) \citet{Storm74}, (27)
\citet{Acke04}, (28) \citet{Lada}, (29) \citet{Terranegra}, (30)
\citet{Warren}, (31) \citet{Herbig}, (32) \citet{Tuthill}, (33)
\citet{Testi}, (34) \citet{Lorenzetti}, (35) \citet{Shevchenko},
(36) \citet{Strom72}, (37) \citet{Ancker00}, (38) \citet{Drew}, (39)
\citet{Malbet06}, (40) \citet{Levreault}, (41) \citet{Berrilli},
(42) \citet{Ancker04}, (43) \citet{Claria}, (44) \citet{Hartmann96},
(45) \citet{MS97}, (46) \citet{Sandell}, (47) \citet{Malbet05}, (48)
\citet{Malbet98}, (49) \citet{MG06}, (50) \citet{Straizys}, (51)
\citet{Muzerolle03}, (52) \citet{White} }
 \tablenotetext{a}{Only minimum values are known.}
 \tablenotetext{b}{\citet{Ancker98} suggest 316$^{+272}_{-103}L_\odot$, but we adopt
the most recent estimate of 100$\pm 35L_\odot$ by \citet{skew}.}
 \tablenotetext{c}{\citet{Herbig} suggest 700pc, but we use $\sim$340pc by
\citet{Tuthill} who derived this distance from the analysis of
the binary companion's apparent motion (luminosity scales with the distance).}
 \tablenotetext{d}{Significant differences in luminosity reported by different authors.
We adopt values from \citet{Drew}.}
 \tablenotetext{e}{The listed luminosities of FU Ori objects are
dominated by accretion luminosity.}

\end{deluxetable}

{\it LkH$\alpha$ 101}: Unlike other high-L objects, LkH$\alpha$ 101
shows a much smaller visibility with the scaled baseline, smaller even
than low-L objects. This indicates a larger NIR emitting area than in
other YSOs. This object was one of the first YSOs imaged with the NIR
interferometry \citep{Tuthill01} and it played an important role in
establishing the existence of the inner disk clearing \citep{Tuthill}.
Prior to the advent of NIR interferometry, the canonical model of the
inner protoplanetary disk was a power-law accretion disk model. Images
of LkH$\alpha$ 101 reconstructed from aperture masking interferometry
data showed clearly that this object has a central clearing in the
disk \citep{Tuthill}. The clearing was attributed to dust sublimation
and LkH$\alpha$ 101 became a prototype example of this new disk
concept.  Our scaled visibilities show that this object differs from
other YSOs studied to date, although the existence of inner clearing
has been confirmed in other objects. It is possible that this
object is indeed more similar to the low-L objects than high-L
objects. There is a large uncertainty in the luminosity, distance and
evolutionary status of this object. Notice, however, that our scaled
baseline in equation \ref{Bscaled} is not affected by luminosity changes
caused by changes in measured distance. There has to be an intrinsic error
in luminosity to change the scaled baseline. For the most recent review
of this object see \citet{Herbig}.

{\it MWC 614}: This objects shows a flat visibility, which means it is
completely resolved at all used baselines. This indicates a much
larger structure than other low-L objects. \citet{skew} suggest the
possibility of a companion star at about 1$''$ distance, which would
lead to an overestimate of the size of the emission region.

\begin{figure}
 \begin{center}
 \includegraphics[width=6.5cm,angle=-90]{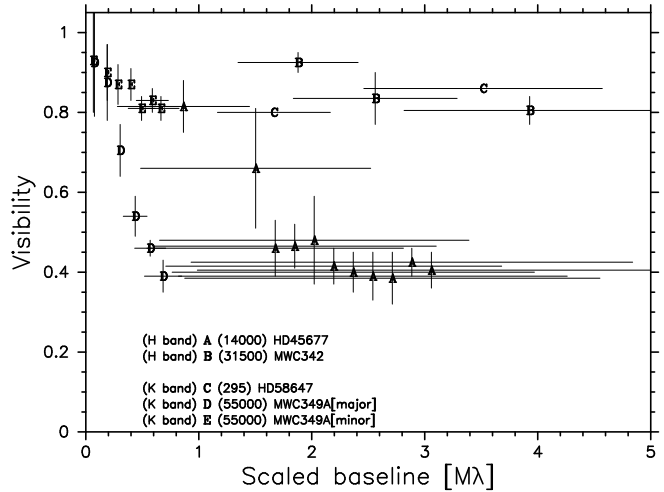}
 \caption{\label{Be_stars}
Objects that are most probably B[e] stars with infrared excess
and erroneously classified as YSOs in observational campaigns.
This figure shows how they also differ from YSOs in their scaled visibility
(see Figures \ref{H-band-data} and \ref{K-band-data}). Images of MWC349 A are
reconstructed from aperture masking interferometry and here we show the visibility
along its major and minor image axes.
}
 \end{center}
\end{figure}

We excluded from this analysis objects that are most probably B[e]
supergiants (MWC 349A: \citet{Danchi,Hofmann}; MWC 342: \citet{Tolya};
HD45677: \citet{skew,deWinter}; HD58647:
\citet{interfer05,Manoj}). Their basic stellar parameters are shown in
Table \ref{table-Be-data} and their scaled visibilities in Figure
\ref{Be_stars}. When compared with YSOs, these objects do not fit into
the clustering scheme recognized in Figures \ref{H-band-data} and
\ref{K-band-data}.

\section{Theoretical interpretations}
\label{section_theory_models}

Clustering of scaled visibility indicates very similar images, hence
presumably similar dust geometry.  Variations in the model parameters
should explain the cluster location, size and shape. The advantage of
modeling a cluster instead of individual objects is that here we can
address the common properties of these types of objects without being
misled by potential peculiarities of individual objects.  We apply
several different types of image models and discuss their constrains
on the dust geometry based on fits to the data clusters.

\subsection{Uniform brightness ring}
\label{Uniform_ring}

A uniform brightness ring emerged as a prototype image model of NIR
interferometry data. It assumes emission from a constant temperature
disk with a central hole, which gives a ring of constant surface
brightness. The inner radius of the ring is attributed to the dust
sublimation, while the outer radius is derived from the total ring
area required by the measured NIR photometric flux.  Physical
justification for this image configuration was found in the puffed up
inner disk model \citep{DDN}. The most recent NIR interferometry
measurements of the image departures from centrosymmetry \citep{skew}
do not support the original version of the puffed up model, but prefer
its more recent derivative by \citet{Isella}.

The visibility function for a uniform brightness ring (along its major
or minor axis when inclined) combined with an unresolved star is (see
\citet{MG01} or \citet{Eisner04} for details)
 \[
 V^{\rm ring}_\lambda(B) = f^\star_\lambda + 2\,\,{1-f^\star_\lambda
 \over \theta_2^2 - \theta_1^2} \left(\theta_2^2\,\,
  {{\bf J}_1(\pi \theta_2 B(i))\over \pi \theta_2 B(i)}\,\, -  \right.
 \]
 \eq{ \label{unif_ring_visi}
  \left.
 - \theta_1^2 \,\,{{\bf J}_1(\pi \theta_1 B(i))\over \pi \theta_1 B(i)}
 \right)
 }
 \eq{\label{Bi}
   B(i)=
   \cases{ B         &   major image axis \cr
                                     \cr
           B\cos (i) &   minor image axis \cr
    }
 }
 where ${\bf J}_1$ is the Bessel function and
the variables are: $f^\star_\lambda$ = fractional contribution of
the stellar component to the total observed flux at the given NIR
wavelength $\lambda$, $i$ = ring inclination angle, and $\theta_1$,
$\theta_2$ = inner and outer ring size.

\begin{deluxetable}{lccc}
\tabletypesize{\scriptsize}
\tablecaption{\label{table-Be-data} B[${\rm e}$] stars}
\tablewidth{0pt}
\tablehead{
\colhead{Object} &
\colhead{Distance} &
\colhead{Luminosity} &
\colhead{Visibility} \\
\colhead{} &
\colhead{(pc)} &
\colhead{($L_\odot$)} &
\colhead{Reference}
}

\startdata
HD58647   & 295$\pm$65 (1)   & 295$\pm$50 (2)          & 2\\
HD45677   & 1000$\pm$500 (3) & 14000$\pm$7000 (4)      & 4\\
MWC349 A  & 1200 (5)         & 55000$\pm$25000 (5,6)   & 6,7 \\
MWC342    & 1000 (8)         & 31500$\pm$16500 (4,8)   & 4\\
\enddata

\tablecomments{
References: (1) \citet{Ancker98}, (2) \citet{interfer05},
(3) \citet{deWinter}, (4) \citet{skew},
(5) \citet{Cohen}, (6) \citet{Danchi}, (7) \citet{Hofmann},
(8) \citet{Tolya}
}
\end{deluxetable}

\begin{figure*}
 \begin{center}
 \includegraphics[width=11.5cm,angle=-90]{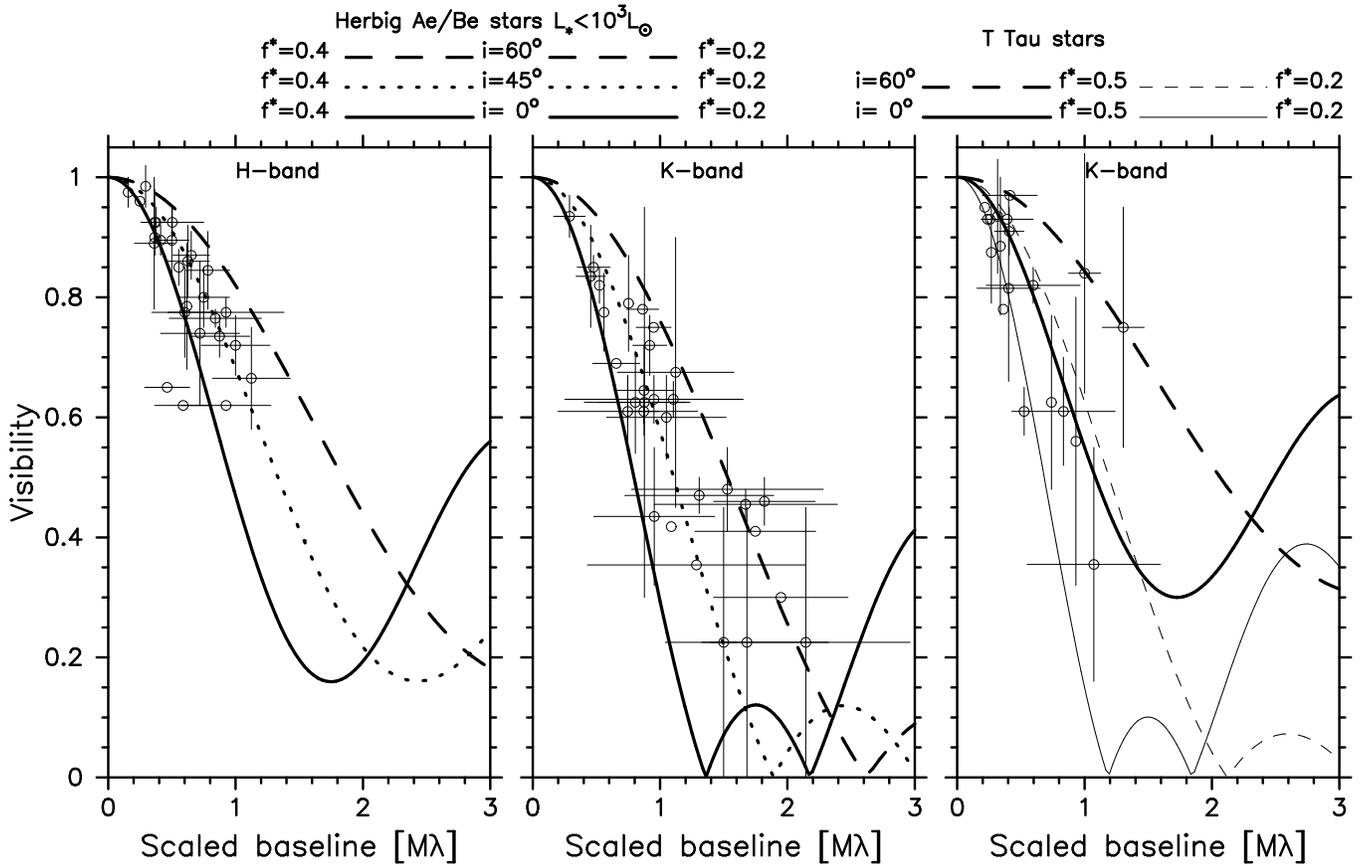}
 \caption{\label{fig_inclinations}
Visibility models of inclined uniform brightness ring. The data is the same as in
Figures \ref{H-band-data} and \ref{K-band-data}. The models assume the ring
temperature $T_D=$1500K and $\Psi=2$ in equation \ref{inner_size} and $T_\star$=10000K
for Herbig Ae/Be stars and $T_\star$=5000K for T Tauri stars.
The fraction of the stellar flux $f^\star$ is set to typical values:
0.4 in H- and 0.2 in K-band for low-L Herbig Ae/Be stars and 0.5 in K-band for
T Tauri stars. While these values reproduce the shape and spread of the low-L
clusters, T Tau stars appear larger than the model. An unrealistically
low value of $f^\star=0.2$ is needed for T Tau stars.
}
 \end{center}
\end{figure*}

\begin{figure}[t]
 \begin{center}
 \includegraphics[width=8.0cm,angle=-90]{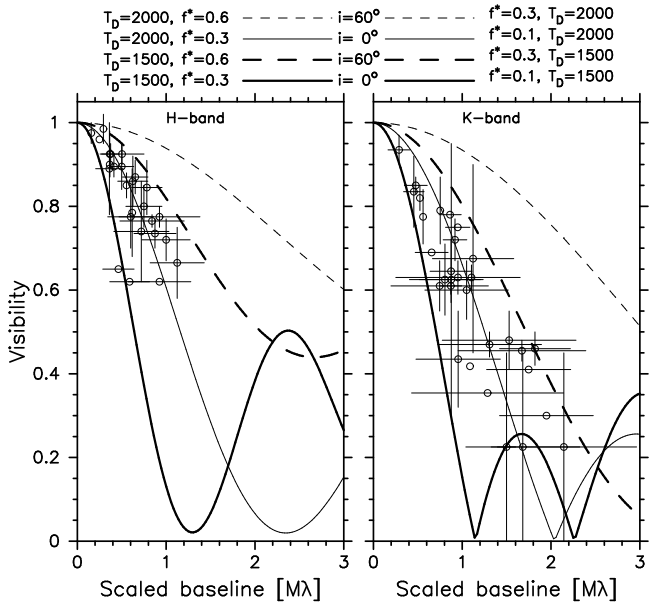}
 \caption{\label{fig_f_Td}
Low-L Herbig Ae/Be stars modeled with inclined uniform brightness
ring.  The model parameters are the same as in
Figure \ref{fig_inclinations}, except for the stellar fraction $f^\star$ and ring
temperature $T_D$. The stellar flux fraction covers the lower and
upper limit, combined with the lower and upper inclination angle
limits from Figure \ref{fig_inclinations}. Data clusters appear smaller
than the range of these parameters and with a clear preference for dust
temperatures of 1500K. T Tau K-band cluster cannot be successfully fitted
with this model (see Figure \ref{fig_inclinations}).
}
 \end{center}
\end{figure}

The inner ring radius $\theta_1$ is controlled by dust sublimation
and grain size. Cooling of smaller grains is less efficient than
bigger grains, hence bigger grains survive closer to the star.
\citet{Vinkovic} has shown analytically that in optically thick
disks made of a mixture of grain sizes, the inner radius is dictated
by the largest grains in the mix. They provide shielding of smaller
grains from direct stellar radiation, which allows smaller grains to
move closer to the star. The inner disk radius of optically thick
disks is, therefore, controlled by the largest grains, that is, by
their sublimation. The most efficient cooling in NIR is archived by
big (micron size or larger) grains because of their gray opacity in
NIR and shorter wavelengths. For that reason, big grains reach the
minimal possible dust distance from the star. Mid-infrared
spectroscopy of YSOs shows that the presence of big grains is a
typical feature of circumstellar dust in these objects
\citep[e.g.][]{Boekel03,Przygodda,Boekel05}. Thus, the inner disk
radius in YSOs is uniquely defined by gray dust grains.

On the other hand, dust sublimation does not provide a unique
solution to the inner radius of protoplanetary disks because we lack
constrains on the exact density structure of these disks. Dust
dynamics, growth and sublimation can make the most inner part of the
disk vertically optically thin and allow more efficient cooling of
the disk. This reduces the local diffuse radiation and enables the
optically thin zone of the inner disk to extend much closer to the
star than the optically thick part of the disk. Hence, the inner
disk radius due to gray dust of temperature $T_D$ is
\citep{Vinkovic}
 \eq{\label{inner_radius}
  R_{in} = 0.0344 \Psi \left({1500K\over T_D}\right)^2
      \sqrt{{L_\star\over L_\odot}} \left[\hbox{AU}\right],
}
 where $\Psi$ depends on the details of the disk structure and
radiative transfer and can be as low as $\Psi\sim 1.2$ for optically
thin inner disks or the maximum of $\Psi=2$ in the case of entirely
optically thick disks. In objects with a
non-negligible accretion, $L_\star$ should be replaced with
$L_\star+L_{\rm acc}$. The scaled baseline (equation \ref{Bscaled}) is
introduced from equation \ref{inner_radius} by considering the disk
size at 1pc and 1L$_\odot$

 \eq{\label{inner_size}
 \theta_1 = 0.0688\Psi \left({1500K\over T_D }\right)^2 {\rm [arcsec]}.
}
 The stellar size $\theta_\star$ follows directly from $L_\star$,
while the outer disk size $\theta_2$ is derived from the requirement
of the total flux being a sum of the stellar and ring component
(i.e., $V_\lambda(B=0)=1$)
 \eq{\label{outer_size}
 \theta_2^2 = \theta_1^2 +{1-f^\star_\lambda \over f^\star_\lambda \cos (i)}
 \cdot {{\mathfrak B}_\lambda(T_\star) \over  {\mathfrak B}_\lambda(T_D)}
 \theta_\star^2,
 }
where ${\mathfrak B}_\lambda$ is the Planck function and $T_\star$
is the stellar temperature. Now we can explore the range of
parameter values constrained by the visibility clusters. We do not
discuss this model for high-L objects because no satisfactory fit is
possible.

\begin{figure}[t]
 \begin{center}
 \includegraphics[width=8.0cm,angle=-90]{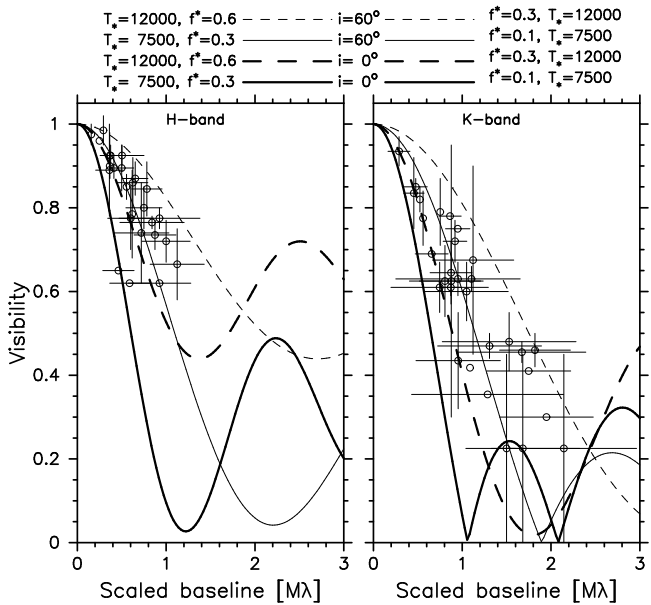}
 \caption{\label{fig_Tstar}
Same as Figure \ref{fig_f_Td}, but for different stellar temperatures. The
disk temperature is set to 1500K.
}
 \end{center}
\end{figure}

\subsubsection{Inclination}

The clusters are expected to spread in size because of randomization of
observed disk inclinations. With equation \ref{unif_ring_visi} we address
only the major and minor axes, which are the image extremes, while the
orientation of baselines can exist in between these two position
angles. Nonetheless, observations are usually performed within a range
of position angles, which increases the chance of covering the image
extremes. These scans of position angles are usually incorporated in the
visibility vertical error bars.

We apply equation \ref{unif_ring_visi} to the low-L Herbig Ae/Be
cluster and T Tauri cluster. We use the dust sublimation temperature
of 1500K and the optically thick disk of $\Psi=2$. The stellar
temperatures are 10000K for Herbig stars and 5000K for T Tau stars.
The fraction of stellar flux $f^\star$ at different wavelengths
depends on spectral shapes of stellar and diffuse radiation. Instead
of letting this be a free parameter, we use observed spectral energy
distributions (SEDs) of YSOs to constrain its value. This way we
actually fit the observed diffuse flux level (relative to the stellar
flux level), which is needed for realistic modeling of both the
circumstellar geometry and the SEDs. Observations show that the
fraction of stellar flux in low-L Herbig Ae/Be stars is in the range
$f^\star\sim$0.2-0.6 in H-band and $f^\star\sim$0.1-0.3 in K-band
\citep{MG01,VIJE06}, while for classical T Tauri stars it is
$f^\star\sim$0.3-0.7 in K-band \citep{Cieza}. Therefore, we choose
medium values of 0.4, 0.2, and 0.5, respectively.

Comparison of models with the data is shown in Figure
\ref{fig_inclinations}. Herbig Ae/Be clusters are nicely reproduced
with this model for the range of inclination angles between 0{\deg}
and 60\deg. The model fails in the T Tau cluster. These objects
appear larger than expected from the model and only unrealistically
small (i.e., inconsistent with observations) values of
$f^\star$=0.1-0.2 can reproduce the cluster size. This is consistent
with the findings of \citet{Akeson05a}, who performed a similar
study and concluded that some T Tau stars appear bigger than the
expected dust sublimation radius. There is also no positive
correlation between size and accretion, just on the contrary, the
opposite trend (decreasing size with increasing contribution of
accretion luminosity) has been reported by \citet{Akeson05a}. Hence,
we conclude that this model cannot provide a satisfactory
explanation of the T Tau data cluster, and we continue with only
addressing model parameters in low-L Herbig Ae/Be stars.

Interestingly enough, the inclination angles up to 60{\deg} are
sufficient to explain the low-L cluster size even though we have not
yet varied other parameters. Since we expect other parameters to
spread the model further, this might be an indication of images
appearing as if they had inclination angles in a smaller range than
expected from random orientations. We address this issue further
below.

\subsubsection{Stellar flux fraction and dust sublimation}
\label{subsubsec_fstar_Td}

Next we explore the above mentioned limits in the stellar flux
fraction. Figure \ref{fig_f_Td} shows models with limiting values of
$f^\star$ combined with disk inclinations. Combining low $f^\star$
with low $i$ and high $f^\star$ with high $i$ brings up extremes in
visibility functions. Comparison with the data shows that clusters
appear slightly more compact than these limits, but this could be
just the result of a small dataset.  However, freedom in choosing
the disk temperature (i.e., dust sublimation temperature) results in
large model deviations from the data, as shown in Figure
\ref{fig_f_Td}. The temperature of 2000K is often suggested to
explain the data in some individual objects, but when it comes to
explaining the collective dataset, then the temperature of 1500K
emerges as the most plausible choice, assuming that visibility data
clusters are indeed a result of very similar physical conditions in
all these objects.

\subsubsection{Stellar temperature and disk opacity}

Objects comprising the data clusters have a range of stellar
temperature from 7500K to 12000K for low-L Herbig Ae/Be stars.  Figure
\ref{fig_Tstar} shows how this spread in temperature affects the
model. Since the stellar spectrum at these temperatures peaks at
wavelengths shorter than NIR, changes in the stellar temperature do
not have a significant effect on the NIR images. Differences of these
models from the 10000K models in Figure \ref{fig_f_Td} are small and
conclusions from \S\ref{subsubsec_fstar_Td} also hold here.

\begin{figure}[t]
 \begin{center}
 \includegraphics[width=8.0cm,angle=-90]{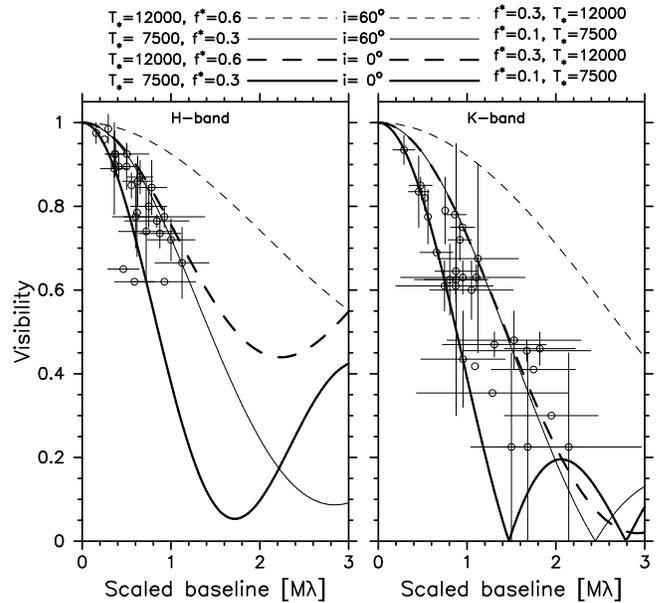}
 \caption{\label{fig_Tstar_thin}
Same as Figure \ref{fig_f_Td}, but now for optically thin disks with
$\Psi=1.2$. Although the model of an optically thin disk would work for small
inclination angles, the data clusters clearly support an optically thick
disk when a full range of inclination angles is considered.
}
 \end{center}
\end{figure}

In addition, we explore the possibility of an optically thin inner disk
and apply $\Psi=1.2$ to these models. Figure \ref{fig_Tstar_thin}
shows that optically thin disks would be an option if disks are almost
face on, but not when the whole cluster of visibility data is
considered.  Since $\Psi{\sim}1.2$ is the minimal possible value,
maybe some intermediate ($1.2\la\Psi<2$) values exist. However, the
NIR excess in Herbig Ae/Be stars cannot be explained by optically
thin disks alone. Instead, an additional circumstellar structure, such
as a puffed up disk or a dusty halo, has to be invoked to reproduce the
observed amounts of NIR flux \citep{VIJE06}.

\begin{figure*}
 \begin{center}
 \includegraphics[width=11.5cm,angle=-90]{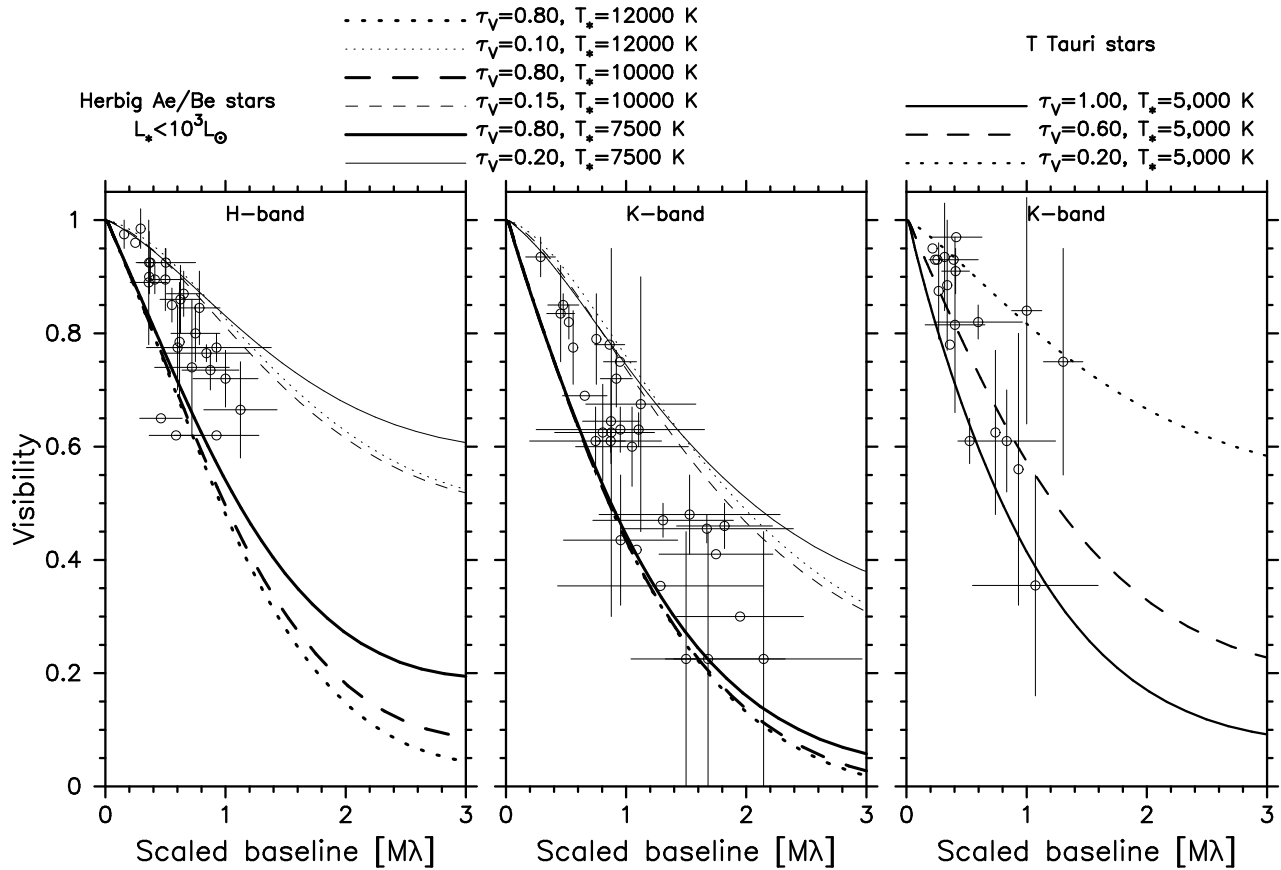}
 \caption{\label{fig_halos}
Visibility models of dusty halos. The models assume a spherical halo
for simplicity. The dust is 1$\mu$m silicate grains sublimating at
1500K. Lines show visibility functions for various optical depths and
stellar temperatures. Optically thin halos are sufficient to explain
the data clusters.
}
 \end{center}
\end{figure*}

\begin{figure}[t]
 \begin{center}
 \includegraphics[width=8.0cm,angle=-90]{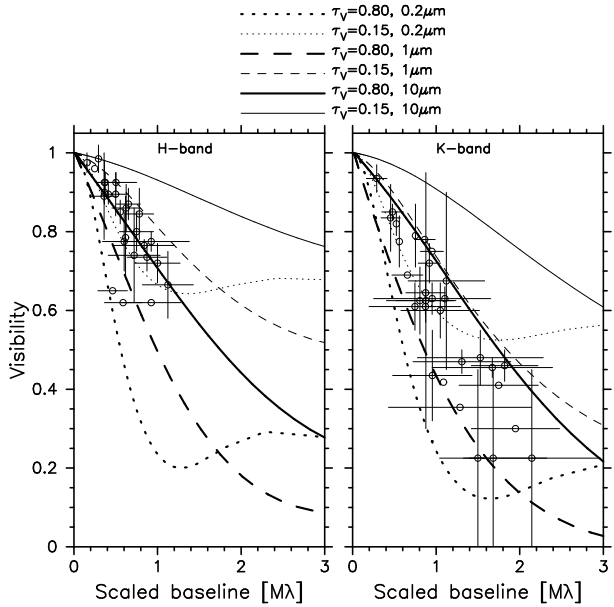}
 \caption{\label{fig_halo_grainsize}
Same as Figure \ref{fig_halos}, but now for different grain sizes.
The data clusters belong to low-L Herbig Ae/Be stars.  The model
stellar temperature is 10000K. Lines show variations in optical depth
and grain size. If dusty outflows exist in these objects, then the visibility
data suggest dust grain size of $\sim 1\mu m$.
}
 \end{center}
\end{figure}

\subsection{Dusty halo}
\label{Subsection_halo}

An alternative model to disk geometries is a dusty halo around the
disk inner regions. Compact ($\sim$10 AU) optically thin halos have
been invoked to explain the NIR spectrum of Herbig Ae/Be stars and
their NIR visibility \citep{Johns-Krull,VIJE06}. The exact physical
mechanism for creating such a halo is not known, but the variability
of YSOs (discussed in \S\ref{subsection_halos_plausible}) supports its
existence as a dusty outflow.

In its most rudimentary form, the halo can be approximated with a
spherical geometry. We emphasize that this is just a convenient
approximation, while a more realistic description of a dusty outflow
probably requires a flattened and clumpy halo.  Nevertheless,
optically thin halos are dominated by direct stellar heating and,
therefore, will maintain similar temperature profiles, no matter the
exact geometry. We can also ignore the disk heating of the halo
because optically thin halos are transparent to the disk emission. The
exact images of the disk and halo model ultimately depend on detailed
properties of the halo and the intrinsic ratio between the disk and
halo surface brightness. These images also depend on the inclination
angle, which is something that we cannot address here with our simple
approximation. We caution that halo models cannot be dismissed by
postulating that they should produce centro-symmetric images
\citep[e.g.][]{Isella06}. Speckle interferometry images of the Herbig
Be star R Mon are an illustrative example of complicated asymmetric
images produced by a parabolic dusty outflow combined with the
inclined disk \citep{RMon}.

The halo radiative transfer is solved with the code DUSTY
\citep{DUSTY}, which takes advantage of the scaling properties of the
radiative transfer problem for dust absorption, emission and
scattering \citep{IE97}. The stellar spectra used in our modeling are
taken from Kurucz models. The silicate dust optical constants are from
\citet{Dorschner95} ($x$ = 0.4 olivine). The radial density profile is
described with $r^{-2}$ since \citet{VIJE06} showed that it can
reproduce the NIR SEDs of Herbig Ae/Be stars. The outer halo radius
has to be large enough to extended beyond the distance where the halo
brightness drops below interferometric or photometric
detection. Hence, we fix it to be 100 sublimation radii. Visibility
functions at specified wavelengths are part of the computational
output from the code. The output visibility already has the stellar
and diffuse components included, thus, we do not need to specify
$f^\star$ in this modeling. The variable that indirectly controls the
amount of diffuse flux in optically thin halos is the radial optical
depth \citep[see Appendix in][]{VIME03}.

The provided spatial frequency $q_{\rm dusty}$ from DUSTY is the
frequency scaled with the angular size of the inner cavity
\citep{IE96}. Since the inner cavity radius $r_1$ for $L_\star=10^4
L_\odot$ is also provided in the output, we can derive the scaled
baseline from
 \eq{
  B_{\rm scaled} = 1.54\times 10^{14} q_{\rm dusty}
  \left({r_1\over {\rm cm} }\right)^{-1}\left({\lambda\over {\rm \mu m}} \right).
 }

Comparison between halo visibility models and data clusters is
presented in Figure \ref{fig_halos}. Dust grains used in the models
are 1$\mu$m in size and sublimating at 1500K. We use stellar
temperatures typical for these stars and find that optically thin
halos can explain the data clusters.  The required visual optical
depth in low-L Herbig Ae/Be stars is $\tau_V\sim 0.15-0.8$, which is
in agreement with $\tau_V\ga 0.2$ derived from fits to SEDs
\citep{VIJE06}. The T Tau cluster displays a slightly larger upper
optical depth limit of $\tau_V\sim 1$.

\begin{figure*}
 \begin{center}
 \includegraphics[width=14.0cm,angle=-90]{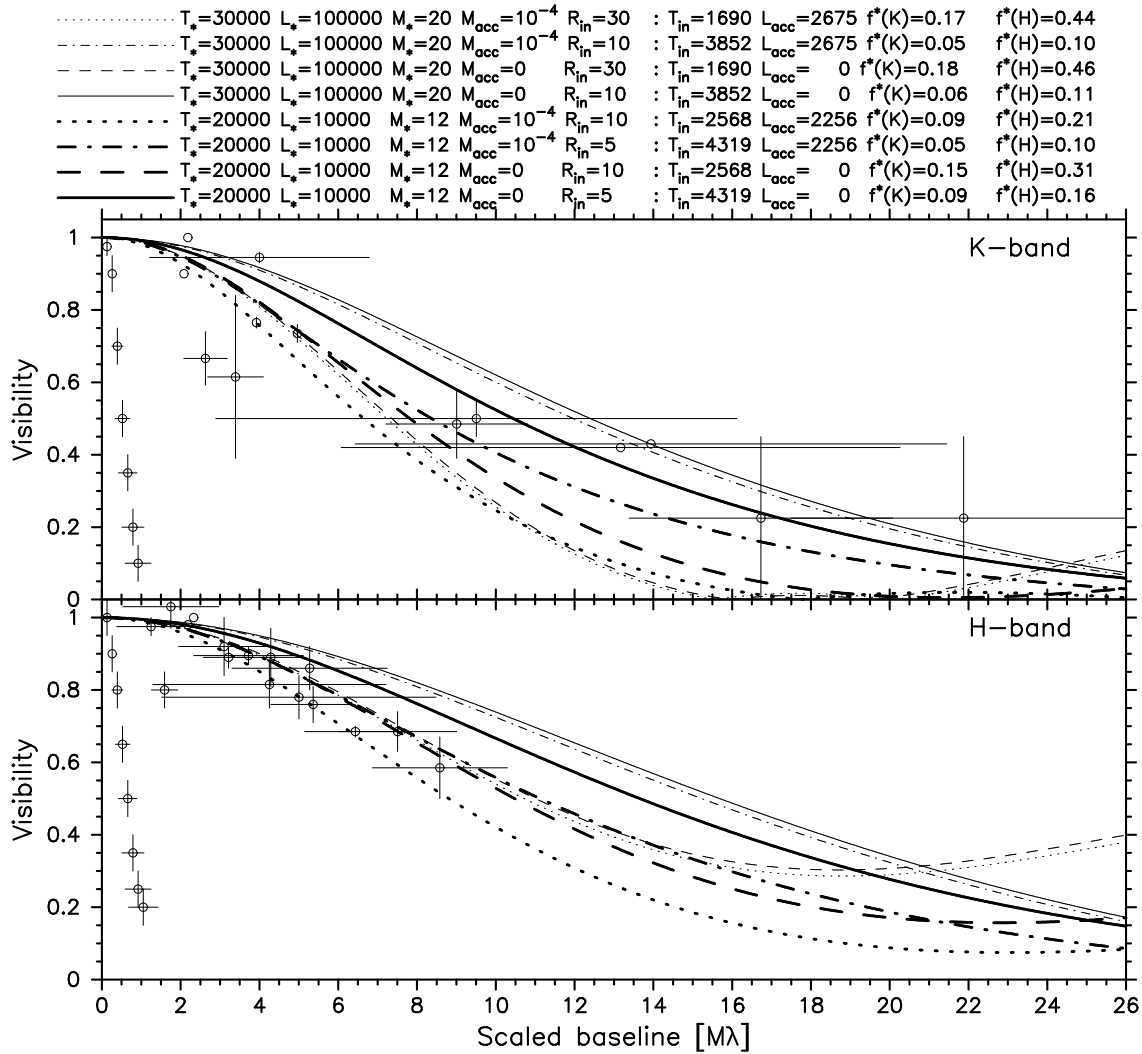}
 \caption{\label{fig_acc_disk_B}
Accretion disk models of high-L Herbig Be stars. The data are the same
as in Figures \ref{H-band-data} and \ref{K-band-data}. The legend
lists parameter values used in modeling: stellar temperature $T_\star$
in K, stellar luminosity $L_\star$ and mass $M_\star$ in solar units,
accretion rate \Mdot in $M_\odot/yr$, and inner disk radius $R_{\rm
in}$ in solar units. Presented models are for face on disks. Other
listed parameters are derived from the model: disk temperature $T_{\rm
in}$ in K at the inner disk edge, accretion luminosity $L_{\rm in}$ in
solar units, and stellar flux fraction in K-band ($f^\star(K)$) and
H-band ($f^\star(H)$). See \S\ref{Accretion_disks} for more details.
}
 \end{center}
\end{figure*}

\begin{figure}
 \begin{center}
 \includegraphics[width=9.0cm,angle=-90]{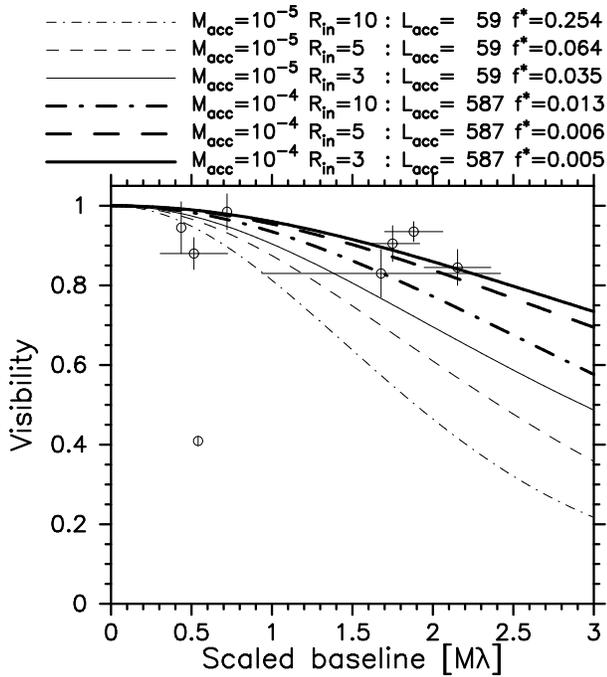}
 \caption{\label{fig_acc_disk_FUOri}
Same as Figure \ref{fig_acc_disk_B}, but for FU Ori stars. The
K-band data are shown. The stellar temperature, mass and luminosity
are 5000K, 1$M_\odot$ and 4$L_\odot$, respectively. Lower than
expected visibilities are most probably an indication of over-resolved
large dusty halos \citep{MG06}.
}
 \end{center}
\end{figure}

The grain size of $\sim 1\mu m$ seems optimal to fit the data.
Figure \ref{fig_halo_grainsize} shows how the visibility changes
when the grain size is reduced to 0.2$\mu$m or increased to
10$\mu$m. These changes are caused by variations in the sublimation
radius $R_{in}$ of the halo, where larger grains reduce and smaller
grains increase this radius. Similar grain sizes fitting the data
mean that the inner halo radii are also very similar.

The model is unsuccessful in explaining the visibility of high-L
Herbig Be stars. The only way to reach the observed visibility
levels of these stars is to keep the halo at tiny optical depths,
which creates a flat visibility due to an unresolved star at
observational baselines. This was the approach used by \citet{VIJE06}
in fitting one high-L star (MWC 297). Since observed visibilities show
a non-flat structure, this model is unsatisfactory. This shows that if
the halo model is correct in low-L objects, then high-L objects differ
from low-L objects by not having a dusty halo, which is not surprising
considering that large luminosities would impose a large radiation
pressure on optically thin dusty halos. Such a scenario would be in
agreement with observations showing that YSOs of spectral type earlier
than about B5 disperse their circumstellar environment much faster
than later spectral types \citep{Fuente98}. Hence, the NIR
visibilities of high-L objects imply that these are stars that
are still surrounded by circumstellar gas and dust, but the process of
dispersal has already started within their immediate environment.

\subsection{Accretion disks}
\label{Accretion_disks}

Prior to the discovery of inner disk holes in low-L Herbig Ae/Be stars
produced by dust sublimation \citep{interfer}, accretion disks were
assumed to be the main source of the NIR excess in YSOs. Since
accretion disks extend closer to the star than the dust sublimation
radius, they are still a favored explanation for the NIR
visibility data of high-L Herbig Be stars
\citep{Malbet06,interfer05,Eisner04,Eisner03,interfer,MG01}.

\subsubsection{Accretion disk model}

The total luminosity of stars with accretion disks is a sum of the stellar
and the accretion luminosities, $L_{\rm tot}=L_\star + L_{\rm acc}$, where
$L_{\rm acc} = G M_\star \Mdot/2R_\star$. Combining stellar heating
\citep{Friedjung} (for $R\ga 2R_\star$) and viscous heating \citep{BellPringle} of
a geometrically thin, optically thick disk results in the disk temperature
\[
T(\theta) = \left[2.19\times 10^{-7}\left({R_\star \over R_\odot}\right)^3
T_\star^4 + 4.21\times 10^{16} \left({M_\star \over M_\odot}\right) \cdot \right.
\]
\eq{
\left. \cdot
\left({\Mdot \over M_\odot/{\rm yr}}\right)
\left(1-\sqrt{\theta_{\rm in}\over\theta} \right) \right]^{1/4}
\left({L_{\rm tot} \over L_\odot}\right)^{-3/8}
\theta^{-3/4},
}
where $\theta$ is the {\it scaled} angular size derived from the disk
radius $R$ and the total luminosity
\eq{ \theta = 2 \sqrt{L_\odot \over L_{\rm tot}} {R\over 1{\rm AU}}.
}

The inner disk edge size $\theta_{\rm in}$ (radius $R_{\rm in}$)
and disk inclination angle
$i$ are free parameters. The stellar fraction of NIR flux
$f^\star_\lambda$ can be computed by integrating the disk emission
\eq{
 {1\over f^\star_\lambda} -1 = {2\cos i\over \theta_\star^2 \,\, {\mathfrak B}_\lambda (T_\star)}
 \int\limits_{\theta_1}^{\theta_{out}} {\mathfrak B}_\lambda (T(\theta)) \theta d\theta,
} where the upper integral limit can be any size where the disk
temperature drops below detection at NIR wavelengths.

The visibility is a sum of narrow rings over the disk surface. Each
ring has the visibility described by equation \ref{unif_ring_visi},
hence the accretion disk visibility is
 \[
 V^{\rm acc}_\lambda(B) = f^\star_\lambda \left[ 1+2\cos (i)\sum\limits_{\theta_1}^{\theta_{out}}
   {{\mathfrak B}_\lambda(T(\theta)) \over  {\mathfrak B}_\lambda(T_\star)} \cdot
   \right.
 \]
 \eq{ \label{acc_disk_visi}
 \left.
   \cdot \left({(\theta+\Delta\theta)^2\over \theta_\star^2} \cdot
   {{\bf J}_1(\pi (\theta+\Delta\theta) B(i))\over \pi (\theta+\Delta\theta) B(i)}
 - {\theta^2\over \theta_\star^2}
 \cdot {{\bf J}_1(\pi \theta B(i))\over \pi \theta B(i)}
 \right) \right],
 }
where $\Delta\theta$ is a small angular step and $B(i)$ is the
scaled baseline defined in equation \ref{Bi}.

Unlike in dust ring models, here we do not have a firm theoretical
limit on the inner radius $R_{\rm in}$, but if the disk accretes onto
the star, then the inner radius of several stellar radii is
expected. The main problem in constraining these models is that
precise stellar parameters are not well known for high-L Herbig Be
stars. On top of that, accretion luminosities are relatively small
compared to stellar luminosities, which makes accretion rates
difficult to constrain.

All these uncertainties are illustrated in Figure \ref{fig_acc_disk_B}
where we compare various accretion disk models with the data (we
ignore anomalously low visibilities of LkH$\alpha$ 101 in this
discussion, as described in \S\ref{section_data}). We display models
with no accretion (a purely reprocessing disk) and with a very high
accretion rate of 10$^{-4}$M$_\odot/yr$ to illustrate how models with
almost any accretion rate can be build to fit the visibility data
\citep[for further discussion see e.g.][]{Malbet06}.  Models are
calculated only for face on inclination, since any loss of flux in
inclined disks can be compensated by an increased inner disk radius
\citep{MG01}. Also, inclination increases the visibility, which again
calls for a larger inner radius to fit the data. Hence, presented
models put a lower limit on $R_{\rm in}$.

\subsubsection{The nature of disk in high-L Herbig Be stars}
\label{HerbigBe_disk_nature}

It seems that some models with $R_{\rm in}\sim 5R_\star$ can fit the
visibility, which would be in agreement with the magnetic accretion
radii implied from T Tau stars \citep{Kenyon}. However, Figure
\ref{fig_acc_disk_B} shows that models with \hbox{${\ga}10R_\star$} are also
applicable. Moreover, a detailed modeling of MWC 297 yields $R_{\rm
in}\sim 17R_\star$ \citep{Malbet06}, which raises the question of what
is the physical process that truncates the disk at these radii.

\citet{interfer05} suggest that the NIR visibility of high-L objects
is due to dust in the accretion disk. In that case the inner radius is
dictated by the dust sublimation temperature. Figure
\ref{fig_acc_disk_B} shows that sublimation temperatures would require
$R_{\rm in}\ga20$. Such disks would be optically thick in their purely
gaseous part inside of the dusty disk hole. This would allow shielding
of the dust from direct stellar radiation, so that the dust can move
closer to the star than the ordinary dust sublimation distance. The
extent of this shielding can be estimated by scaling the high-L
visibility clusters in Figures \ref{H-band-data} and \ref{K-band-data}
to the location of low-L clusters. We find that all the stellar flux
shorter than $\sim 0.6\mu m$ has to be removed to make this process
work. The obtained scaling factor is $\sim$7, which also shows how
smaller high-L objects are from the dust sublimation size typical of
low-L objects.

Optically thin gas can be optically thick in some molecular and
atomic lines, but it would be transparent at other wavelengths and
incapable of producing such a severe reduction of the stellar flux.
The shielding is, therefore, plausible only if the gas is completely
optically thick, which also implies high accretion rates ($\ga
10^{-7}M_\odot/yr$) \citep{Muzerolle04,Bell,Hartmann93}. On the other
hand, this creates a problem because the side effect is a strong
near IR radiation that can actually overshine the dusty disk.
\citet{Muzerolle04} demonstrate this in the case of a puffed up dusty
disk, where gas emission from accretion disks of $\Mdot\ga
10^{-6}M_\odot/yr$ would dominate the NIR. Since shielded dusty
accretion disks do not have such a puffing, their near IR dust
emission is smaller than from puffed up disks and, therefore,
competes with smaller accretion rates.

The effect is even more pronounced in NIR visibilities where a
bright optically thick gas would affect images by decreasing the
observed inner disk radius. Comparison between disk models of
different inner radii in Figure \ref{fig_acc_disk_B} shows that any
such decrease would significantly alter the visibilities. Thus, if
dust is the source of NIR visibility of high-L objects, then the
gas inside of the dust sublimation radius has to be optically thick
for stellar radiation, but not bright in the NIR at the same
time. Scattering of UV by low-density gas was suggested by
\citet{interfer} as a possible solution, but this cannot work because
{\it i}) large optical depths are needed to block almost all of the
scattered stellar flux, and {\it ii}) even if all of the stellar UV
flux is removed, it is still not enough to make high-L scaled
visibilities similar to low-L visibilities. This dilemma of how to
reconcile such properties of the gaseous disk within the dust
sublimation zone is actually quite old, and it was already recognized
in a slightly different context by \citet{Hartmann93}.

It is, therefore, possible that instead of dust, the NIR flux of
high-L objects originates from gas emission despite confusions about
the inner disk radius. This would suggest that these stars have high
accretion rates and their circumstellar environment is evolutionary
similar to FU Ori objects (young counterparts of T Tau stars). Since
both types of objects are of a similar age, they probably share
similarities in the accretion process. Fast pre-main-sequence
evolution of massive stars actually requires high accretion rates
\citep{Palla00}. Also, observed high mass loss rates imply high
accretion rates in massive stars \citep{Cesaroni,Shepherd}, similar
to massive winds supported by high accretion rates in FU Ori stars
\citep{Sandell,Calvet}.  The main difference is the ability of
highly luminous Herbig Be stars to relatively quickly disperse their
circumstellar matter \citep{Fuente98}. 

The standard viscous disk model predicts proportionality between the
disk mass and the accretion rate \citep[e.g.][]{Calvet2000}. Thus, high
accretion rates would suggest higher disk masses in high-L than in
low-L objects. Observations, however, show exactly the opposite
\citep{Fuente03}. It may be that the dispersion of circumstellar
matter around high-L stars starts with disk erosion due to disk and
stellar winds. This may reduce the disk mass and explain the observed
decrease in disk masses. Their destructive nature may also explain why
these stars form a data cluster at $\sim$7 times larger scaled baselines
than low-L YSOs instead of producing a range of values due to disk
optical depth variations between objects. Namely, it may be that only
high density and high accretion disks can survive so close to these
luminous stars, which would result in disks with very similar NIR
signatures.

High accretion rates may also be implied from the recently observed
correlation between the accretion rate and the square of the stellar
mass of low-L YSOs \citep[e.g.][and references
therein]{Calvet2004,Muzerolle05,GarciaLopez,Natta06}. If this
correlation is not just a selection effect \citep{Clarke}, then it
probably extends to high-L objects, too. The required accretion rates
would be $\Mdot\ga 10^{-7}M_\odot/yr$, which is exactly what we
suggest based on the NIR visibilities.

There are additional complications in deducing the nature of disks
around Herbig Be stars. MWC 166 is, for example, a binary with a
mean distance of $\sim$1.7AU between components and an orbital
eccentricity of e=0.28 \citep{Pogodin}. The binary is surrounded by a
common gaseous envelope and the existence of the disk in this system
is a more puzzling problem than our general discussion about Herbig Be
stars. We also ignore any contribution from the NIR free-free
emission. Although this is a good approximation for Herbig Be stars,
it is possible that some objects have a significant amount of
free-free emission \citep[e.g MWC 297:][]{Porter}

\subsubsection{FU Ori accretion disks}

Accretion disks in FU Ori stars are corroborated with NIR
visibility measurements, as evident from Figure
\ref{fig_acc_disk_FUOri} where we compare the visibility data of FU
Ori stars and accretion disk models. Modeling is simplified in this
case due to similarities in stellar properties of FU Ori stars. We use
face-on disks and stellar temperature, mass and luminosity of 5000K,
1$M_\odot$ and 4$L_\odot$, respectively. Classical accretion disks of
$R_{in}$=3-5$R_\star$ and $\Mdot=10^{-4}-10^{-5}M_\odot/yr$ reach the upper
observed visibility levels \citep[see also][]{Malbet05,Malbet98,MG06}.
There is a problem with anomalously low visibility of Z CMa-SE, which
is more consistent with T Tau visibilities. Other FU Ori stars also
show a slightly smaller visibility than expected from pure accretion
disks, which indicates an additional larger completely resolved
circumstellar structure.  \citet{MG06} argue that this is due to a
large dusty envelope. Such a disk+envelope structure was previously
predicted from the overall infrared spectral energy distribution
\citep{Hartmann96}. This configuration and derived accretion rates
were recently confirmed by modeling the infrared spectra of FU Ori
stars taken with the Spitzer Space Telescope \citep{Green}. However,
\citet{Quanz} modeled their mid IR visibilities of FU Ori without any
additional structure such as a dusty envelope and concluded that
the presence of accretion disk is sufficient to explain this object.

\section{Discussion}
\label{section_doscussion}

\subsection{The importance of dust dynamics}
\label{subsection_dust_dynamics}

We usually approach this kind of study with the assumption that
circumstellar environments of YSOs share enough similarities to be
successfully described with one universal theoretical model. This is a
very strong assumption considering that the spectra reveal very
dynamic gaseous disks around YSOs \citep[e.g.][]{Mora04}.  Since
gravity and gas drag are two forces that dictate dust dynamics within
protoplanetary disks, dust should also display a very dynamic
behavior. The infrared spectra provide evidence that this is indeed
the case: the mid IR dust features exhibit clear signatures of small
grains in the disk surface \citep{Acke04}, which is possible only
if dust dynamics persistently replenishes the surface with these
grains \citep{Dullemond05}.

Dust dynamics is, however, ignored when it comes to formulating dusty
disk models used in calculating synthetic infrared spectra and
images. The disk is modeled with a smooth surface that seemingly
appears dynamically passive. This is a good approximation when it
comes to modeling data taken at one epoch, but multi-epoch data
reveals that YSOs are far more complicated than such simple disk
models. Basically all pre-main-sequence objects are spectrally and
photometrically variable, differing only in the amplitude and rate of
variability. This variability includes also the near and mid IR
wavelengths where dust emission dominates \citep[e.g.][]{ChenJura,
Grinin00, Herbst_variable, Skrutskie, Liu, Hutchinson, Prusti}.

While some mid IR variability can be attributed to variations in the
stellar (or accretion) heating of the dust, many of these
variabilities are {\it not} accompanied by changes in the luminosity
of the central source. Such cases are a clear sign of dust dynamics
producing modulations in the dust emission and/or stellar
obscuration. On top of that, \citet{Vinkovic} showed analytically
that the dust sublimation zone of optically thick dusty disks cannot
be constrained purely by radiative transfer. Sublimation leads to
vertical optical thinning of the disk, which, combined with dust
dynamics, leads to entirely new radiative transfer solutions. This
result demonstrates that inferring a realistic geometry of dust
distribution in the inner disk region of optically thick
protoplanetary disks requires dust dynamics as well as radiative
transfer.

Although the NIR visibilities are currently observed with a limited
u-v coverage, dust dynamics responsible for strong variabilities
should be detectable in multi-epoch visibility observations. In
addition, the newly employed NIR closure-phase measurements enable
interferometric detections of dusty disk inhomogeneities, as
recently demonstrated in the case of AB Aur \citep{MG_ABAur}. Time
evolution of such brightness asymmetries may significantly improve
our understanding of processes responsible for the infrared
variability of YSOs.

This is an important topic because current disk models have problems
in explaining variabilities caused by transient dust obscurations
of the central star (UXOR variables). This variability is manifested
as changes in dust extinction, accompanied by the ``blueing'' effect
and increased polarization at minima \citep{Shakhovskoj, Rodgers,
Rostopchina, Grinin01, Grinin00, Skrutskie, Grinin94, Hutchinson}. The
photometric minima effects are a result of increased relative
contribution of scattered starlight by dust in the total spectrum
\citep{Natta00}.

Temporal properties of obscuration events affecting visual and NIR
wavelengths indicate that dust clouds usually appear in the inner
disk regions at or close to the dust sublimation zone. If they
belong to inhomogeneities {\it constrained to the disk} then disk
inclinations have to be large \citep{Dullemond03}. The necessesity
for such a correlation was originally suggested by \citet{Natta97}.
If we assume the most optimistic inner disk puffing of $H/R\sim$0.2
\citep{VIJE06} then the line of sight would be affected by the inner
disk when inclination angles are $i\ga$80\deg. This is inconsistent
with the imaging data, which show disks with smaller inclination
angles. In particular, the UXOR star CQ Tau \citep{Shakhovskoj} has
$i=48\deg^{+3}_{-4}$ derived from the NIR visiblities
\citep{Eisner04}, $i=33\deg\pm 5$ from the mid IR imaging
\citep{Doucet} and $i\sim 60-70\deg$ from mm-imaging
\citep{Testi03}, while the UXOR star VV Ser has $i=42\deg^{+6}_{-2}$
derived from the NIR visiblities \citep{Eisner04}.

We do not agree with the interpretation by \citet{Isella06} that
their derived disk inclinations (from fits to the NIR visibility) of
40-55{\deg} for CQ Tau and 50-70{\deg} for VV Ser are in agreement
with the dust obscuration model. These inclinations would actually
support the idea of dust clouds being ejected to $\ga$9 (CQ Tau) and
$\ga$4.5 (VV Ser) vertical scale heights above the dusty disk, which
would dynamically decouple them from the disk (i.e., their dynamics
would not be controlled by gas-drag any more).

\subsection{Are dusty halos a plausible option?}
\label{subsection_halos_plausible}

Dust obscuration events are easier to explain if we allow for a
possibility of dust clouds being ejected to moderate polar angles
above the disk mid-plane. Notice that this would still imply a dependence
of variability on the inclination angle, which can explain the
observed weak correlation between inclinations measured in $v\sin i$
and the amplitude of photometric variability \citep{GrininKozlova00}.
Similar correlation between the observed polarization and $v\sin i$
does not exist, but polarimetry supports the idea of a flattened halo
above the disk \citep{Yudin}. In addition, numerical models reproduce
photometric and polarimetric variability of UXORs using the ad hoc
assumption of obscuring dust clouds above the scattering dusty disk
\citep{Natta00}. We note that the variable accretion luminosity model
put forward by \citet{Herbst_variable} in order to explain the
variability of UXOR objects was successfully challenged by the
proponents of dust obscuration model
\citep{Rostopchina,Grinin01,Grinin00}.

Once a cloud is out of the dense gaseous disk, gas drag force becomes
negligible and radiation pressure takes over. The cloud would be
eventually blown away, creating a dusty halo-like outflow above the
disk. Changes in such an outflow would manifest itself by changes in
the near and mid IR thermal emission. This IR variability is not
dependent on the inclination angle, assuming optically thin
emission. It would also be uncorrelated with the stellar spectrum
variability. Although simultaneous visual and infrared observations
are rare, such variability events have been documented
\citep{ChenJura,Eiroa02,Prusti,Hutchinson}.  Particularly interesting
is the case of HD 163296 in which \citet{Sitko} detected an event of a
major increase in the near and mid IR emission. Since HST imaging
shows a Herbig-Haro flow from this star \citep{Grady00}, they conclude
that the observed infrared ``flare'' is possibly an ejection of a
forthcoming Herbig-Haro object. If correct, this would directly link
stellar outflow with the infrared variability due to dusty outflow.

Unfortunately, despite these seemingly convincing arguments, the
existence of a dusty outflow is not straightforward and obvious. The
biggest drawback in halo models is the lack of a known force capable
of lifting a dust cloud out of the disk. The common assumption in
circumstellar disk modeling is that gravity and gas drag do not
allow dust to exit the high density gaseous disk. Hence, the
existing studies of NIR visibilities are dominated by disk models.
Circularly symmetric images are sometimes invoked as a feature of
dusty outflows not in agreement with NIR interferometry
observations. However, in \S\ref{Subsection_halo} we argue that this
is an ill-formed argument because halos can certainly display
asymmetric images.

There was great excitement for some time about the possibility of
cometary activity in Herbig Ae/Be stars detected in spectral line
variations. This is called the $\beta$ Pictoris phenomenon, after the same
effect first observed in the dust debris disk of $\beta$ Pic
\citep{Grady_BetaPic_review}. According to this interpretation,
accretion episodes accompanied by redshifted absorption components in
a number of metal lines are signatures of infalling evaporating
comets. This would nicely fit into the above scheme of transient dust
clouds. Unfortunately, further observations showed that the
composition of infalling gas is consistent with the accretion disk gas
\citep{Natta_betapic}, while numerical models revealed that spectral
signatures of comets cannot be detectable due to suppression by
strong stellar winds in Herbig Ae/Be stars \citep{Beust}. Since then
there have been no other suggestions for a process that could lift a
dust cloud out of the disk.

A new imperative to explore dusty halos may come from the NIR
visibility studies. The large NIR excess (often called the NIR
``bump'') of Herbig Ae/Be stars requires a vertical disk puffing at
the inner disk edge \citep{Isella06,DDN}. The NIR images based on this
model have a large skewness - the brightness ratio between two opposite
sides of the image centered on the star. \citet{skew} recently
measured the NIR interferometry closure phases that can reveal the
amount of skewness and found that observations do not support model
predictions\footnote{\citet{skew} suggest that curved puffed up rim
model by \citet{Isella} may be more suitable, but notice that this
model also produces images with a large skewness \citep{Isella06}}.
However, their claim is based mostly on high-L objects that do not
exhibit visibilities consistent with the puffed up disks anyway (see
\S\ref{HerbigBe_disk_nature}). The main question is whether low-L
Herbig Ae/Be stars can be described with this model or not. It turns
out that closure phases are inconclusive for their skewness determination
due to insufficient telescope resolution.

On the other hand, detailed modeling of the NIR visibilities with the
puffed up disk model is not successful in reproducing strong near IR
bumps \citep[MWC 758 and VV Ser:][]{Isella06}. It was exactly these
objects with a strong NIR bump that \citet{VIJE06} point out as the
most puzzling and most difficult to explain. The model also has
problems with reconciling variability of UXOR stars with disk
inclinations derived from the model, as already described
above. Despite these difficulties, the model of curved puffed up rims
by \citet{Isella06} is quite promising and represents the most
advanced description of the inner disk region so far.

\subsection{Dust grain size in the inner disk region}
\label{grain_size_discussion}

Dust properties of the inner protoplanetary disk can also be deduced
from the models of observed near IR visibilities. Since in
\S\ref{HerbigBe_disk_nature} we argue that NIR visibilities of high-L
objects are not dust related, we limit our discussion on dust grain
size to low-L YSOs only.

As already described in \S\ref{Uniform_ring}, the inner radius of
optically thick multigrain disks is controlled by the largest grains
in the mix. Also, dust growth and dynamics may change the opacity
structure of the inner disk and form a large optically thin zone
populated by big ($\ga$1$\mu$m) grains extending closer to the star
than the optically thick part of the disk.  Results from
\S\ref{Uniform_ring} support models of optically thick inner disks
containing big grains. Smaller grains can hide inside the dusty disk,
but we can see only big grains that populate the disk surface and
shield smaller grains from the direct stellar heating.  Dusty halo
models of NIR visibilities also favor micron size grains, but no
larger than $\sim$10$\mu$m (see \S\ref{Subsection_halo}).
Visibilities are, therefore, clearly suggesting that circumstellar
dust exists in the region where submicron grains cannot survive when
directly exposed to the stellar radiation.

Presence of big grains in the inner disk has been suggested before.
\citet{Grady_big_grains} modeled variations in the UV and visual
spectra of UX Ori and obtained the best fit with grains
$\ge$0.15$\mu$m, which suggests considerable grain growth in
comparison with the ISM dust. \citet{Boekel04} obtained spectra from
the \hbox{1-2 AU} zone of the immediate surrounding of several low-L
Herbig Ae/Be stars using mid IR interferometry. These spectra are
dominated by micron size thermally processed (crystalline) dust
grains, which differs from the outer (2-20 AU) region where grains
are smaller ($\sim$0.1$\mu$m) and less crystalline. This is in
agreement with results based on the NIR visibilities, where we
expect such a grain size gradient due to dust sublimation. Small
grains ($\la$0.1$\mu$m) around low-L Herbig Ae/Be stars sublimate at
$\ga$1 AU, while micron size grains can survive within 1 AU. These
big grains populate the inner disk surface and dominate detected
visibilities and spectra of the inner disk.

\subsection{The difference between high-L and low-L YSOs}

The clear distinction between visibilities of low-L and high-L YSOs
may be a sign of more extensive dissimilarities between these two
classes of objects. The change in circumstellar environment detected
by visibilities appears at $\sim$10$^3L_\odot$, which corresponds to
the spectral type of about B3-B5 during the PMS phase. Here we point
out some other significant changes in the YSO properties that happen
around that spectral type.

{\it Dispersal of circumstellar gas}: The mean gas density within a
radius of 0.8 pc around high-L Herbig stars decreases by almost two
orders of magnitude during their fast \citep[$\la 3\times 10^5$
yr:][]{Palla93} evolution to the main sequence \citep{Fuente98}. In
contrast, environments of low-L stars experience a decrease of less
than an order of magnitude. The NIR interferometry targets high-L
objects that have still not dispersed their immediate environment. In
\S\ref{HerbigBe_disk_nature} we speculate that disk accretion rates
under such conditions are high enough to enable a gaseous accretion
disk to dominate the NIR emission and visibilities. This differs from
low-L objects, where disks have time to evolve to the point where
accretion rates become low enough to make the gaseous disk
transparent, allowing dust emission to dominate the NIR.

{\it Disk Mass}: Disk masses in high-L Herbig stars are significantly
lower than in low-L \citep{Fuente03}. The trend is most noticeable in
the ratio of disk to stellar mass, but it is also significant when the
absolute disk mass is considered. It could be that these stars evolve
through the PMS phase too quickly to accrete a massive disk before the
star disperses its circumstellar environment.

{\it Radiation pressure}: Detailed calculations of radiation pressure
force on dust grains are very complicated and uncertain due to our
limited knowledge of geometrical properties of individual dust grains
\citep{Burns}. Nevertheless, we can exploit some general properties of
the ratio of radiation pressure force to gravity, $\beta$, and derive
an estimate for the size of grains repelled by the star ($\beta>1$).
The ratio $\beta$ is generally proportional to
$L_\star/M_\star$. Differences in the shape of stellar spectrum, grain
refractive indexes, and grain density can scatter the value of $\beta$
by an additional order of magnitude \citep{Burns}. But luminosity
differences between low-L and high-L YSOs are so large that, in the
first approximation, a comparison of $\beta$ values in YSOs is
dominated by $L_\star/M_\star$. A PMS star of $L_\star=10^3L_\odot$
has $M_\star\sim 6M_\odot$ \citep{Palla93}, which means $\sim$170
times larger $\beta$ than in the Solar System. High-L objects have
even larger $L_\star/M_\star$, hence, even larger $\beta$. The Sun can
eject dust particles of $\sim$0.2$\mu$m in size
\citep{Landgraf,Burns}, while other particles smaller than
$\sim$10$\mu$m in the Solar System have $\beta\sim 0.01-1$. This means
that we can safely assume that YSOs of $10^3L_\odot$ repel all dust
grains smaller than $\sim$10$\mu$m. In general, high-L YSOs have this
limit at grain sizes larger than 10$\mu$m because their luminosity is
$>10^3L_\odot$. It is difficult to maintain dusty structures around
high-L objects with such high values of $\beta$. A dusty halo would be
very efficiently dispersed. Puffed up disk rims cannot maintain their
vertical stability and would be efficiently blown away. Only dust
within the optically thick gaseous disk or strongly dragged by gas toward
the star can resist the radiation pressure force. The situation is
completely different for low-L objects. A $10^2L_\odot$ Herbig star has
$L_\star/M_\star\sim 30$ \citep{Palla93}, hence the upper grain size
limit of dust with $\beta\ga 1$ is in the micron range. Curiously
enough, this is consistent with smaller grains not surviving in the
surface of the inner disk (see \S\ref{grain_size_discussion} above).

{\it Variability}: The range of photometric variabilities of high-L
YSOs is much smaller ($\Delta$V$\la$1.5 mag) than that of low-L YSOs
($\Delta$V$\la$6 mag)
\citep{RodgersPhD,Herbst_variable,Bibo,Finkenzeller}. This dependence
of variability on luminosity is not understood. It may be that high
luminosities are masking underlying accretion variabilities (unlike FU
Ori stars where the stellar luminosity is much smaller than the
accretion luminosity) or that dust sublimates so far away from the
star that the probability for a transient dust obscuration is highly
reduced \citep{GrininKozlova00}. Near IR visibilities, however,
indicate that differences in the geometry of circumstellar matter
distribution may be the reason. Another interesting property of the
variability data is the correlation between the NIR flux excess
and the range of variability amplitudes. \citet{Herbst_variable} show
this for Herbig Ae/Be stars by plotting the variability range in V
band against the ratio of NIR excess luminosity to the stellar
luminosity.  \citet{Skrutskie} show this for T Tau stars by plotting
variations of intrinsic (K-L) color excess against the amplitude of
the K-band variability. If the puffed-up inner dusty-disk rim model is
responsible for V-band variability (which we challenge; see
\S\ref{subsection_dust_dynamics}) then this correlation indicates
temporal variations in the rim's height. The halo model, on the other
hand, explains this correlation by connecting the amount of dust in
the halo with the dynamics of dusty outflow. A larger NIR flux
requires a larger optical depth of the halo, which in turn is a result
of enhanced dust dynamics supplying the halo with dust. Side effects
of this enhanced dynamics are dust obscuration events and NIR
emission variabilities. Since the existence of such a halo is
consistent only with low-L YSOs (see \S\ref{Subsection_halo}), the
variability amplitude is smaller in high-L YSOs because they lack the
halo and its dust dynamics. Unfortunately, the current data are not
sufficient for establishing which theory is correct.

\section{Summary}
\label{section_summary}

\begin{figure}[t]
 \begin{center}
 \includegraphics[width=9.0cm]{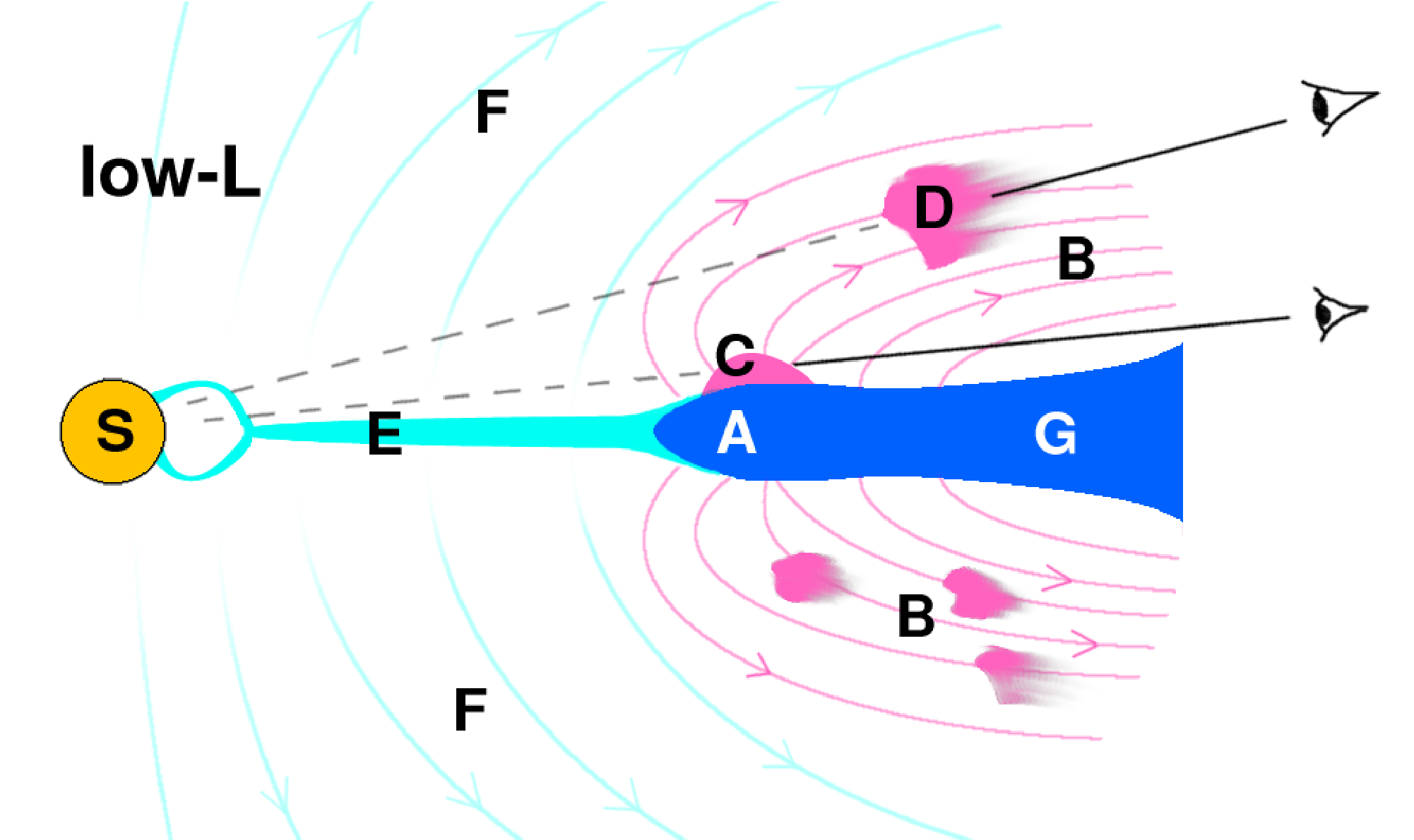}\\
~\\
 \includegraphics[width=9.0cm]{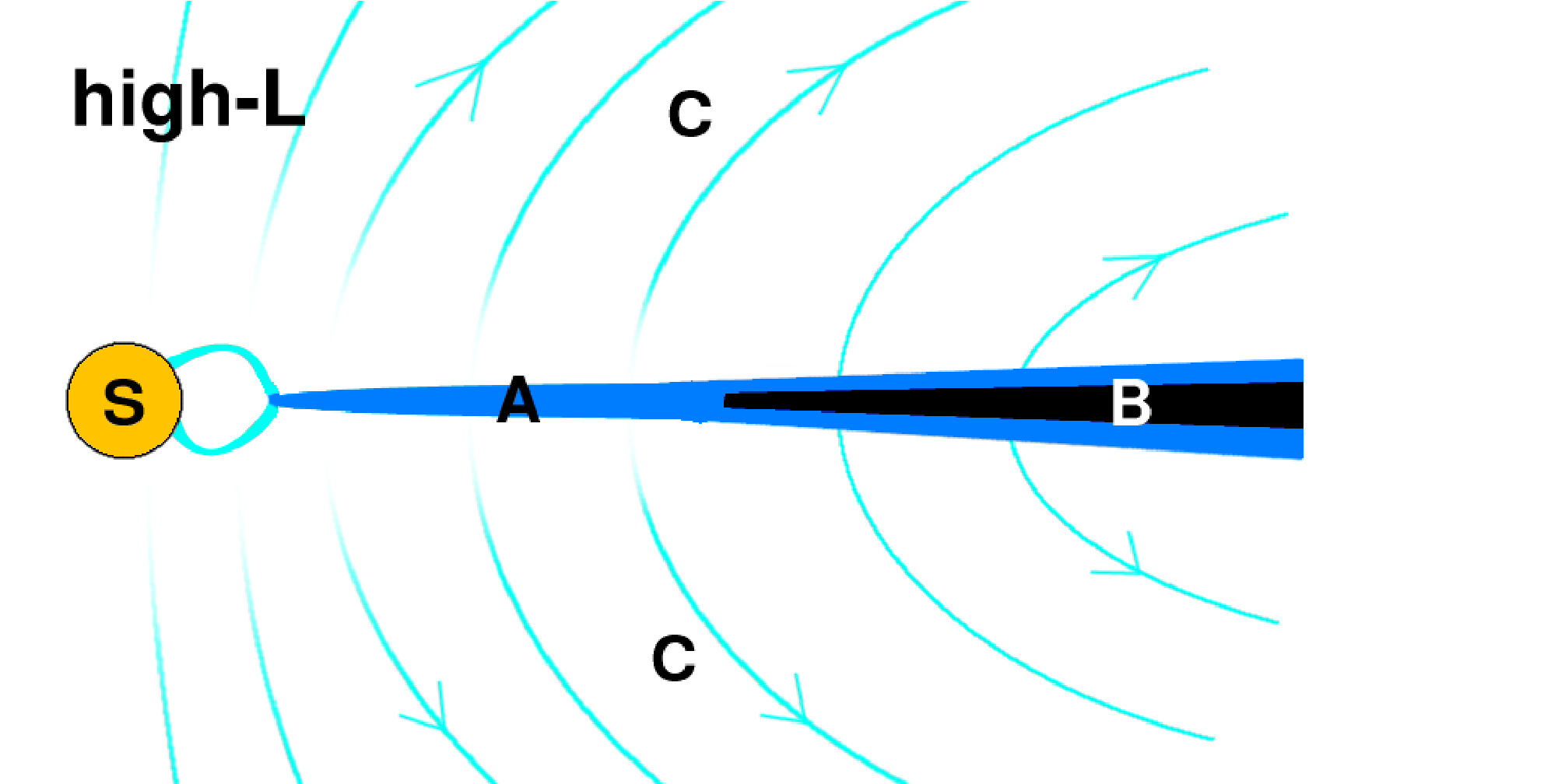}
 \caption{\label{sketch}
The emerging picture of the inner protoplanetary disk structure
based on observed properties of high-L and low-L YSOs.  {\it Top}:
Low-L objects have two competing models explaining their NIR
visibilities and anomalously high NIR excess: (A) puffed-up inner
disk rim and (B) dusty outflow creating a halo around the inner
disk. Detected variability due to dust obscuration events suggests that either
(C) the height of puffed-up rim is variable and temporarily blocks the
view toward the star or (D) clumps of dust appear in the dusty outflow
and occasionally intercept the line of sight. The disk inside the zone
of dust sublimation (E) is optically thin, while the rest is optically
thick (G) due to dust.  Gaseous stellar and disk wind (F) are also
present.  {\it Bottom}: High-L YSOs have a simpler structure. The star
is surrounded by optically thick gaseous accretion disk (A), which
extends much closer to the star than the dust sublimation
distance. But dust still may survive within the optically thick disk
interior (B). Intense gaseous stellar and disk wind (C), combined with
the stellar radiation pressure, are efficiently dispersing the
surrounding environment.  (S) marks the star.
}
 \end{center}
\end{figure}

We have reviewed all published NIR (H- and K-band) visibility data
of YSOs and devised a method for their model-independent comparison.
The method is based on scaling the distance of objects and their
luminosity out of the measured baseline (equation \ref{Bscaled}).
This removes the apparent dependence of the object's size on: {\it i})
its distance and {\it ii}) radiative transfer scaling due to
luminosity. Hence, the visibility dependence on scaled baseline
detects inherent differences in the geometry of circumstellar matter
distribution without applying any additional theoretical
model-dependent assumption.

The comparison shows a clear distinction between low-L YSOs
($L_\star\la 10^3L_\odot$) and high-L YSOs ($L_\star\ga 10^3L_\odot$),
as already suggested by previous studies
\citep{interfer05,Eisner04}. Low-L visibilities cluster at
spatial scales $\sim$7 times larger than scales derived from the
visibility clustering of high-L YSOs. Next, we analyze the observed
visibility clusters with three types of image models. Modeling the
whole data cluster instead of individual objects reveals or
reaffirms some collective properties of these objects.

The first model is the uniform brightness ring, where we use dust
sublimation as the boundary condition for the ring's inner
radius. High-L YSOs are inconsistent with dust sublimation and appear
much too small. The model was also not successful in explaining the
size of T Tau cluster. These stars appear slightly larger than model
predictions. The model was more successful in low-L Herbig Ae/Be stars where
the visibility cluster can be modeled with optically thick rings of
0-60{\deg} inclination and a dust sublimation temperature of
$\sim$1500K. The second model is the optically thin dusty halo. It
explains the T Tau cluster with halos of $\sim$0.2-1.0 visual optical
depth. Low-L Herbig Ae/Be stars require optical depths of
$\sim$0.15-0.8. Halos made of micron size grains provide the best fit
to these data clusters. Finally, the third model is a classical
accretion disk. We use this model on high-L Herbig Be stars and show
that it can accommodate observed visibilities. The model does not
constrain the accretion rate because equally successful fits can be
built with $\Mdot=0-10^{-4}M_\odot/yr$. We argue, however, that the
NIR emission from accretion disks must be due to gas and not dust,
hence, accretion rates must be high in order to produce required
thermal emission from the gas.

We also discuss variability properties of YSOs and made an attempt to
incorporate them into the existing models of inner protoplanetary
disk.  We argue that dust obscuration events detected through
photometric and spectroscopic variability are caused by dust clumps
too high above the dense inner protoplanetary disk to be considered
dynamically part of the disk. Instead, we advocate a model where dust
is ejected out of the disk and blown away by the radiation pressure.
This would create a clumpy dusty halo-like outflow above the inner
disk. Unfortunately, we are not aware of any physical process that
could lift the dust out of the disk. Without this process the only
alternative is to assume that the inner disk rim is puffed-up and
undergoes occasional height variations responsible for dust
obscuration events. But this model is also problematic because it
requires disk inclinations lager than actually observed in objects
with dust obscuration events. All these facets of evidently
complicated circumstellar environment of YSOs are sketched in Figure
\ref{sketch}.  Grain size also plays a very important role in this
environment. We argued for large ($\ga$1$\mu$m) grains in the surface
of inner disk region of low-L YSOs based on theoretical and
observational evidence.

In addition to the difference in their visibility clusters, high-L and
low-L YSOs differ in many other observational aspects. High-L Herbig
Be stars: {\it i}) are much more efficient in dispersal of their
circumstellar gas and dust, {\it ii}) have smaller disk masses, {\it
iii}) have strong radiation pressure capable of destroying
circumstellar dust structures made of grains up to 10$\mu$m or more in
size, {\it iv}) are far less photometrically variable. All these
properties clearly indicate that the geometry of circumstellar gas and
dust distribution around high-L YSOs is fundamentally different from
those in their lower luminosity counterparts. Our current
understanding of these environments is far from satisfactory and
it is largely incomplete.

\acknowledgments

We thank Doron Chelouche and Anatloy Miroshnichenko for useful
discussions. We also thank the anonymous referee for useful comments
that have improved this paper.  D.V. acknowledges supports by the NSF
grant PHY-0503584 and the W.M. Keck Foundation. T.J. acknowledges the
hospitality and financial support of the Institute for Advanced Study
during his visit to the institute.

\end{document}